\documentclass[11pt]{article}
\usepackage[pdftex]{graphicx}
\usepackage{hyperref}
\usepackage{cite}
\usepackage{bm}
\usepackage{amsmath}
\usepackage{dsfont}
\usepackage{amsmath,amsfonts,amssymb,amsthm,nccmath,latexsym,mathtools}
\usepackage{xcolor}
\usepackage{amsmath}
\usepackage{cancel} 
\usepackage[normalem]{ulem} 
\usepackage{amssymb}
\usepackage{amsbsy} 
\usepackage{soul}

\usepackage{footnote}

\hypersetup{
	colorlinks   = true, 
	urlcolor     = blue, 
	linkcolor    = blue, 
	citecolor   = red 
}

\def\no{\nonumber}

\def\d{\delta}

\def\p{\partial}

\def\na{\nabla}

\def\th{\theta}
\def\k{\kappa}

\def\t{\tilde}

\def\ll{\lambda_{(l)}}

\def\L{\mathcal{L}}

\def\be{\begin{equation}}
\def\ee{\end{equation}}
\def\ba{\begin{align}}
\def\ea{\end{align}}

\def\H{\mathcal{H}}
\def\mg{\sqrt{-g}}

\def\A{\mathcal{A}}
\def\B{\mathcal{B}}
\def\D{\mathcal{D}}

\def\V{\mathcal{V}}

\def\P{\mathcal{P}}

\def\sptmr{(\mathcal{M}, \textbf{g}, \bm{\nabla})}
\def\hech{\mathcal{H}}

\def\fr{\frac{1}{2}}

\def\sptqr{(\bnl, \bm{q}, \bm{\twod})}

\def\ll{\lambda_{(l)}}

\def\mt{T^{(m)}}

\def\m4{\mathcal{O}(\epsilon^4)}

\def\bml{\bm{{l}}}
\def\bmk{\bm{{k}}}

\def\bnl{\B^{\rm (null)}}
\def\en{\rho\epsilon^{\rm N}}
\def\ov{\bm{\Omega}\cdot \bm{V}}
\def\twod{~^{(2)}D}
\labelformat{section}{Section #1} 
\labelformat{subsection}{Section #1} 
\labelformat{subsubsection}{Section #1}
\labelformat{subsubsubsection}{Section #1}
\labelformat{equation}{Eq.~(#1)} 
\labelformat{figure}{Fig.~#1} 
\labelformat{subfigure}{Fig.~\thefigure#1} 
\labelformat{table}{Tab.~#1} 
\labelformat{appendix}{Appendix #1}
\begin{document}
\title{{\bf{\Large A note on gravity and fluid dynamic correspondence  on a null hypersurface}}}
\author{Krishnakanta Bhattacharya\footnote{krishnakanta@dubai.bits-pilani.ac.in}$~^{1}$,  Sumit Dey\footnote{Sumit.Dey@fuw.edu.pl, deysumit92@gmail.com}$~^{2}$ and Bibhas Ranjan Majhi\footnote{bibhas.majhi@iitg.ac.in}$~^{3}$\\
$^{1}${\small{Department of General Science, Birla Institute of Technology and Science, Pilani,}}\\
{\small{Dubai Campus, International Academic City, Dubai, United Arab Emirates}}\\
$^{2}${\small{Faculty of Physics, University of Warsaw, Pasteura 5, 02-093 Warsaw, Poland}}\\
$^{3}${\small{Department of Physics, Indian Institute of Technology Guwahati, Guwahati 781039, Assam, India}}
}
\date{\today}
\maketitle
\begin{abstract}
In the extensive literature on fluid-gravity correspondence formulated on null hypersurfaces, the Carrollian and membrane paradigm approaches have predominantly employed a \textit{timelike foliation}. By contrast, within the \textit{null foliation formalism}, only the momentum conservation law, expressed through the Damour-Navier-Stokes (DNS) equation, has been established. In this work, we revisit the \textit{null foliation formalism} for a generic null hypersurface and extend it to include the energy conservation law, continuity equation, and related relations, all derived from the covariant conservation of an appropriately defined energy-momentum tensor. This development complements the existing literature on the fluid description of gravitational dynamics in the null foliation framework.

\end{abstract}






\section{Introduction}
Numerous signs in the literature point to the possibility that gravitation is not a fundamental force but rather an ``emergent phenomenon''. The concept came from Sakharov's work \cite{Sakharov:1967pk}, and numerous other researchers later supported it \cite {Jacobson:1995ab, Volovik:2000ua, Padmanabhan:2009vy}. These are some of the arguments that support this paradigm: (i) The thermodynamic laws can be directly linked to the field equations of many classes of gravitational theories \cite {Padmanabhan:2002sha, Cai:2005ra, Paranjape:2006ca, Padmanabhan:2006ag, Kothawala:2009kc, Chakraborty:2015aja,Bhattacharya:2022mnb,Dey:2021rke}, (ii) The action functional of various theories can be identified as the free energy of the spacetime \cite{Gibbons:1976ue, Padmanabhan:2004fq, Padmanabhan:2009kr, Kolekar:2011bb,Bhattacharya:2017pqc}, (iii) Obtaining field equations from a thermodynamic extreme principle \cite{Padmanabhan:2007en, Padmanabhan:2007xy}, (iv) Possibility of obtaining the idea of energy equipartition theory for defining the microscopic degrees of freedom \cite{Padmanabhan:2013nxa, Chakraborty:2014rga, Bhattacharya:2024vbx}, (v) Dynamical equations for gravitation manifesting as the Navier-Stokes-like equation for wide class of theories \cite{Damour:1979wya, PhysRevD.18.3598, Gourgoulhon:2005ng, Padmanabhan:2010rp, Chakraborty:2015hna, Bhattacharya:2020wdl}, (vi) Association of every solution of Navier-Stokes (NS) equation in $d$ dimension with a ``dual'' solution of Einstein's equation in $d+1$ dimensions \cite{Bredberg:2011jq} (also see \cite{De:2018zxo,De:2019wok,Dey:2020ogs}) \textit{etc.}  
All of the aforementioned phenomena broadly illustrate how gravitational theories bear striking similarities to two seemingly unrelated pillars of emergent physics: fluid dynamics and thermodynamics. Despite several limitations—most notably, the lack of consensus on the nature of the underlying microscopic degrees of freedom—the thermodynamic interpretation of gravity has proven to be remarkably robust and internally consistent. The same seems to be not so, however, for the fluid-gravity correspondence, especially when extended to generic null surfaces and the broader interpretation of gravity as an emergent phenomenon.

The connection between gravity and fluid dynamics was first uncovered by Damour \cite{Damour:1979wya, PhysRevD.18.3598}, whose work on null surfaces led to the membrane paradigm, treating black hole horizons as fluid-like systems \cite{Price:1986yy}. This analogy matured with the AdS/CFT correspondence, where gravitational perturbations in AdS spacetimes were shown to map onto relativistic hydrodynamics on the boundary \cite{Bhattacharyya:2008kq, Bhattacharyya:2008ji, Bhattacharyya:2007vjd, Policastro:2002se}, culminating in a precise correspondence between incompressible Navier–Stokes solutions and vacuum Einstein solutions \cite{Bredberg:2011jq}. Around the same time, Padmanabhan offered a thermodynamic perspective, demonstrating that Einstein’s equations on null surfaces can be derived from entropy extremization and are equivalent to the Navier–Stokes equations, thereby framing gravity as an emergent thermodynamic phenomenon that also reproduces key results such as the KSS bound \cite{Padmanabhan:2010rp}. 

 A major breakthrough in this direction appeared when is was shown that the gravitational dynamics of a stretched surface, which is located arbitrarily close to an internal null boundary, say the event horizon, is similar to that of the Carrollian fluid \cite{Ciambelli:2020ftk,Freidel:2022vjq,Petkou:2022bmz,Bagchi:2023rwd,Bagchi:2023ysc,Freidel:2024emv,Bagchi:2025vri}. A host of recent works \cite{Ciambelli:2018xat, Ciambelli:2018ojf, Petkou:2022bmz, Freidel:2022bai, Armas:2023dcz,Ciambelli:2023mvj,Donnay:2019jiz,Herfray:2021qmp} speak in favour of this formalism, which is based on stretched horizon- a timelike proxy for the event horizon. Subsequent work in this gave rise to the Carrollian membrane paradigm. In relativistic fluid dynamics, the energy-momentum tensor (EMT) plays a central role, forming the foundation upon which the entire fluid description is constructed. The fundamental conservation laws of energy and momentum emerge naturally from appropriate projections of the EMT conservation equation $\na_aT^{ab}_{\rm (fluid)} = 0$. A comparable EMT structure also appears in the membrane paradigm \cite{Price:1986yy}, and in recent formulations of Carrollian hydrodynamics for the stretched horizon \cite{Freidel:2022vjq, Freidel:2024emv}.

 The gravitational dynamics of an internal null boundary at finite distances in the spacetime $\sptmr$ is naturally associated to fluid dynamics under different paradigms as mentioned above. At any point $p \in \mathcal{M}$ of spacetime, we can have a local approximate Rindler horizon corresponding to an accelerated observer. Such a Rindler horizon is generated by an approximate time-like boost Killing vector field whose norm vanishes on the Rindler horizon. Because the Rindler horizon is a one-way null surface, the Rindler observer can associate local thermodynamic entities associated with the event $p$ of the spacetime. A generic integrable null surface can locally be approximated by a Rindler horizon \cite{Chakraborty:2015aja}. Hence, the dynamics of the null surface can locally be studied by such Rindler observers. The essential difference that arises in the paradigms mentioned above is due to the way the spacetime manifold in the neighborhood of $\hech$ is foliated. In both the membrane paradigm and the Carroll membrane paradigm, the spacetime in the vicinity of the internal null boundary is foliated by a family of timelike hypersurfaces with the null limit leading to $\hech$. However, following Carter \cite{carter1997extended}, there can also be another foliation of $\hech$ by the family of null hypersurfaces. The gravitational dynamics of $\hech$ that need to be studied are independent of such a null foliation.
 In the Carrollian membrane paradigm, that exists in literature, the vicinity of the null boundary $\hech$ (located at a finite distance) is foliated by a family of three-dimensional timelike surfaces $\mathcal{T}$. Such a family of timelike surfaces is defined by the scalar function $r(x^a) = \text{constant} > 0$. The null boundary is obtained in the limit $r(x^a) = 0$. That is, essentially, the null limit to $\hech$ is obtained by taking the limit $r \rightarrow 0$. However, such a family $\mathcal{T}$ is not foliated by spacelike cross-sections/cuts (which would in turn provide the notion of a time evolution vector field $\bm{t}$ and hence predispose the structures $\hech$ and $\mathcal{T}$ with a Galilean picture). Instead, Carroll structures are imposed on these timelike surfaces $\mathcal{T}$ as well as the null boundary $\hech$ \cite{Freidel:2022vjq, Freidel:2024emv}. With such a Carrollian structure implemented on $\mathcal{T}$, an appropriate Carrollian fluid energy-momentum tensor can be described on the stretched horizon \cite{Freidel:2022vjq, Freidel:2024emv}. In the null limit, this Carrollian (surface) energy-momentum tensor matches with the one introduced in \cite{Chandrasekaran:2021hxc} up to an overall negative sign. The (vacuum) Einstein field equations are then shown to be conservation laws of such a Carrollian fluid energy-momentum tensor. The corresponding viscous stress tensor, pressure, energy density, momentum density, and heat current have been derived for such a Carrollian fluid \cite{Freidel:2022vjq, Freidel:2024emv}.


Consequently, a key open question remains: Does there exist a similar fluid description for the paradigm, where the spacetime in the neighborhood of the internal null boundary $\hech$ is foliated by a family of null surfaces as opposed to timelike foliation adopted so far in the Carroll membrane paradigm? Given the significant difference in the construction of foliation, it is worthwhile to analyze the fluid description of the gravitational dynamics starting out with a possible fluid energy-momentum tensor as the usual starting point in the null foliation case. 
On the other hand within the analysis of null surfaces initiated by Damour and further developed by Padmanabhan, it has been shown that a momentum conservation equation—famously known as the Damour-Navier-Stokes (DNS) equation—can be derived from gravitational dynamics. However, unlike in classical fluid dynamics, other fundamental relations such as the continuity equation and the energy conservation law have not yet been established within this framework. Moreover, the whole classical fluid description is related to the conservation of fluid energy-momentum tensor. This gap indicates that the fluid-gravity correspondence, especially in its emergent gravity interpretation for generic null surfaces, remains incomplete. To achieve a more robust and comprehensive description, it is crucial to investigate whether these additional conservation laws can also be derived.

In this paper, we provide a systematic description of fluid-gravity correspondence for a generic null surface. Our work proposes a novel extension of fluid-dynamic techniques that have recently been developed for the stretched horizon \cite{Price:1986yy,Freidel:2022vjq, Freidel:2024emv}, adapting them to the unique characteristics of null surfaces. Unlike the stretched horizon, which retains a timelike nature, a null surface poses significant challenges due to its degenerate geometry and the absence of a well-defined rest frame. By addressing these issues, our approach provides a more comprehensive framework for describing the fluid-like behavior of null surfaces, broadening the scope of hydrodynamic descriptions in gravitational systems. Our argument does not rely merely on finding the similarity between null dynamical equations with the Navier-Stokes equation and identifying the fluid parameters based on analogy. Instead, we follow a conventional approach of fluid dynamics, which is centered around the null Brown-York energy-momentum tensor-- similar to the fact that the fluid-dynamic description is centered around the energy-momentum tensor of the fluid. We show that some of the fluid parameters can be identified naturally from the different components of null BY tensor, which agrees with the previous identification which was based on analogy. Furthermore, embracing the usual route of fluid dynamics, where the fluid equations (such as the energy conservation equation, momentum conservation equation, etc., which are known as the Navier-Stokes equation) can be obtained from the suitable projection of EMT, we show that the entire fluid-gravity description for a generic null surface can be formulated from the conservation of null BY tensor. Using this approach, not only do we obtain the age-old Damour-Navier-Stokes (DNS) equation, but we also show that the other relations, such as the energy conservation relation and the continuity equation can also be obtained. 

 The paper is presented in the following parts. In the following \ref{secprerequisites}, we have discussed the prerequisite knowledge on the part of relativistic fluid dynamics as well as on the geometry of the null surface, which will be useful for subsequent discussions. In \ref{secBYformulation}, we introduce the Brown-York formalism-- first the original one (valid for the timelike surface) and then its recent null extension. In \ref{secGRAVFLUID}, we discuss how the null Brown-York-like tensor provides the description of fluid-gravity correspondence, where we derive the Damour-Navier-Stokes equation, continuity equation as well as the Null-Raychaudhuri equation from the null Brown-York tensor. In \ref{Aside}, we analyze the covariant gravitational dynamics w.r.t. an adapted coordinate system established on $\hech$ and make connection with the non-relativistic fluid setting. Finally, we provide the discussions related to our analysis in \ref{secDISCUSS}.

Let us describe the notations and conventions used. We will be working with the mostly positive $(-,+,+,+)$ signature of the metric and employ the geometrized unit system where $c,
\hbar$ and $G$ have been set to one. The lowercase Roman alphabets $a,b, \cdot \cdot \cdot$ represent the four-dimensional spacetime indices. The spatial indices on any three-dimensional spacelike surface are designated with the lowercase Latin alphabets $\mu, \nu, \cdot \cdot \cdot$. The indices on the null surface will be designated with the lowercase Latin alphabets, with a tilde over them $\t \mu, \t \nu, \cdot \cdot \cdot$. The uppercase Roman alphabets $A, B,\cdot \cdot \cdot$ will be reserved for the coordinates of a spatial two-dimensional cross-section/cut of the null surface. In \ref{secBYformulation}, $\mathbf u_a$ represents arbitrary timelike/spacelike normal (in contrast to $s_a$, which is a spacelike normal). The tensor $\mathbf{h}_{ab}$ represents the induced metric on the corresponding arbitrary (timelike or spacelike) surface defined by $\mathbf u_a$ (whereas $h_{ab}$ is the induced metric of the timelike surface, described by the normal $s_a$). Similarly, the other boldfaced quantities, such as $\mathbf{K}_{ab}$, $\mathbf{T}^a$, $\boldsymbol{\pi}_{ab}$ etc., refer to geometric or physical quantities associated with this arbitrary timelike/spacelike surface described by $\mathbf{h}_{ab}$. In contrast, the non-bold quantities, such as $K_{ab}$, $T^a$, etc., are those associated specifically with the timelike surface whose induced metric is $h_{ab}$. Furthermore, boldface notation is sometimes used to indicate that the corresponding vector or tensor expressions hold irrespective of whether the indices are covariant or contravariant.


\section{Preambles and prerequisites} \label{secprerequisites}

\subsection{Fluid equations}
In the discussion of (relativistic) fluid mechanics, the constitutive relations are obtained from an energy-momentum tensor (EMT) and baryon current. The EMT $T^{ab}$ encapsulates the density and flux of energy and momentum in spacetime, providing a comprehensive description of the distribution and flow of energy and momentum within a fluid. 
%
The components of the energy-momentum tensor $T^{00}_{\rm (fluid)}$, $T^{0\alpha}_{\rm (fluid)}$ and $T^{\mu \nu}_{\rm (fluid)}$ represent energy density, momentum density and stress respectively. Similarly, the baryon current is also conserved which is expressed via the continuity equation.
In relativistic hydrodynamics, there are five independent parameters, three components of fluid velocity (relativistic four components of velocity are constraint by $v_av^a=-1$), pressure, and (rest) mass density, that are required to describe fluid completely. In order to uniquely determine these dynamical variables, we require the relativistic Euler equations, derived from the conservation of the energy-momentum tensor and the continuity equation for the conserved rest-mass density. The conservation of the EMT is given by,
\begin{align}
\na_aT^{ab}_{\rm (fluid)}=0~.
\label{confluid}
\end{align}
The conservation of the EMT \eqref{confluid} yields four equations that describe the evolution of energy and momentum, which are coupled to the pressure, velocity, and density. This can be shown as follows. It is known that if one takes a timelike projection of the conservation law \ref{confluid}, i.e.
\begin{align}
v_b\na_aT^{ab}_{\rm (fluid)}=0~, 
\label{enconfluid}
\end{align}
 it yields the energy conservation relation of the fluid, where $v_b$ is the four-velocity of the fluid. On the other hand, if one considers the spacelike projection of \ref{confluid}, i.e.
 \begin{align}
 \Delta^c_b~ \na_aT^{ab}_{\rm (fluid)}=0~, 
 \label{momconfluid}
 \end{align}
 provides the momentum conservation of the fluid, where $\Delta^a_b=\delta^a_b+v^av_b$ is the projection tensor, which is orthogonal to the four-velocity $v_a$. Note that the above \ref{momconfluid} is a set of three dynamical equations. In addition to that, we have the baryon conservation equation for the baryon current $J^{a}$  given as,
\begin{align}
\nabla_a J^a=0~.\label{coneqn}
\end{align}
To close the system, we require a thermodynamic equation of state.
Thus, \ref{enconfluid}, \ref{momconfluid} and \ref{coneqn} together with the equation of state make a set of five closed equations that are required to determine the five unknown quantities stated earlier. Translating to the non-relativistic case, the solution of the five dynamical variables are enabled via the three components of the Navier-Stokes equation, the mass continuity equation for the fluid density and an equation of state (to close the system) that typically relates pressure to fluid density and internal energy (or temperature). When the equation of state depends on temperature, then an additional equation for heat flow has to be supplemented.

\subsection{A generic null hypersurface}\label{secgeometry}
In the Riemannian spacetime manifold $\sptmr$, equipped with the Levi-Civita connection $\bm{\nabla}$, a null hypersurface is a surface of codimension one such the induced metric on it is degenerate. Since, in our construction, we consider a generic integrable null surface, it can be defined via a scalar function $u(x^a) = 0$. The null normal $\underline{\bml} = l_a \bm{d} x^a$ to such a surface is given by,
\begin{eqnarray}
	l_a = -e^{\hat{\rho}}\p_a u ~,
	\label{norml}
\end{eqnarray}
with $\hat{\rho}$ being a smooth scalar field defined on the null surface $\H$. The negative sign in \eqref{norml} accounts for the fact the null generators $l^a$ to $\H$ are future-directed by an appropriate choice of the scalar function $u(x^a)$. On account of the null nature of the normal to $\H$, $\bml$ cannot be provided a unique normalization since $\bml \cdot \bml = 0$. The fact that we have considered only a single null surface, which is the support of the null generator $\bml$, we cannot take notions of derivatives of quantities away from the null surface $\H$. To circumvent this problem, following Carter \cite{carter1997extended}, we consider not just a single null surface but rather a family of them. That is, we foliate the spacetime $\sptmr$ in the neighborhood of $\H$ by a stack of integrable null surfaces defined by $u(x^a) = c$, where $c$ is a constant. So we have a family of integrable null surfaces $\H_{u}$ out of which $\H_{u=0} = \H$ is our chosen integrable null hypersurface. This construction then extends the support of the null generator $\bml$ to not just over $\H$ but rather over a finite region in the spacetime. {This then facilitates the notion of taking derivatives of quantities away from the null surface}. It is worth mentioning that such a foliation of the ambient spacetime in the neighborhood of $\H$ is non-unique. However, all geometric quantities that will be described for the null surface will be independent of any such foliation that has been introduced. The integrability of the null surfaces implies that it satisfies the Frobenius identity,
\begin{equation}
	\bm{d}\underline{\bm{l}} = \bm{d} \hat{\rho} \wedge \underline{\bm{l}} ~.
	\label{frobeniusdual}
\end{equation} 
The fact, that the null normal lies on $\H$ itself, does not allow the construction of a projection tensor (that can be created out from the ambient spacetime metric and the null normal) onto the null surface. For that, we need to have the notion of a vector field that is transverse to the null hypersurface. But before introducing such a transverse vector field, let us consider a $3+1$ foliation of the null family $u(x^a) = c$ by a stack of spacelike surfaces $\Sigma_t$ defined by the scalar function $t(x^a) = \text{constant}$. The spacelike surfaces are assumed not to intersect with each other. Hence, a local neighborhood of the spacetime manifold in the vicinity of $\H$ can be coordinatized by $(t,x^1,x^2,x^3)$, with $x^\mu = (x^1,x^2,x^3)$ being the spatial coordinates on a given $\Sigma_t$. The time evolution vector field $\bm{t} = \bm{\p_t}$ that connects the same spatial points $x^{\mu}$ of the neighboring slices $\Sigma_t$ satisfies the condition $t^a \p_a t = 1$. The cuts/intersections of the null surface $\H$ with the spacelike familty $\Sigma_t$ is defined as $\B^{\rm (null)} \equiv \H \cap \Sigma_t$ which form a $2$-dimensional submanifold of $\H$. With this, we introduce a unique notion of a transverse auxiliary null vector field $\bmk$, defined as,
\begin{equation}
	\bm{l} \cdot \bm{k} = -1, ~~~ \bm{k} \cdot \bm{k} = 0 ~~ \text{and} ~~ \bm{k} \cdot \bm{e_{{A}}} = 0 ~,
	\label{defnauxiliaryk}
\end{equation}
where $\{\bm{e_{{A}}}\}$ are the set of two spacelike basis vectors on $\B^{\rm (null)}$. The null generators $\bml$ provide a notion of outgoing null vector field whereas the auxiliary null vector field $\bmk$ is ingoing w.r.t the codimension two submanifolds $\B^{\rm (null)}$. 
With the null generators and the auxiliary null vector field, one can define two projection tensors
\begin{align}
	\Pi^a_{~b}=\d^a_b+k^al_b~,
	\no 
	\\
	q^a_b=\d^a_b+l^ak_b+k^al_b~.
\end{align}
Here $\Pi^a_{~b}$ is only a projection tensor (and not an induced metric) with the following properties:
\begin{align}
	\Pi^a_{~b}l^b=l^a~,~~~\Pi^a_{~b}l_a=0~,~~~\Pi^a_{~b}k^b=0~,~~~\Pi^a_{~b}k_a=k_b~,
	\no 
	\\
	\Pi^a_{~b}\Pi^b_{~c}=\Pi^a_{~c}~.~~~~~~~~~~~~~~~~~~~~~~~~~~ \label{PIABCON}
\end{align}
On the contrary, $q^a_b$ is the projection tensor onto the two-surface $\B^{\rm (null)}$ upon which both $l^a$ and $k^a$ are the normals (i.e. cross-section of $\H$). The properties of $q^a_b$ are as follows:
\begin{align}
	\mathbf{q}\cdot \mathbf{l}=\mathbf{q}\cdot \mathbf{k}=0~,
	\no 
	\\
	q^a_bq^b_c=q^a_c~.
\end{align}
Unlike $\Pi^a_{~b}$, $q^a_b$ is the induced metric on $\B^{\rm (null)}$. 
The intrinsic geometry of the null surface is provided by the first fundamental form, which is the metric $\bm{q}$ induced from the ambient metric $\bm{g}$ from the spacetime manifold $\sptmr$. The codimension of two surfaces $\bnl$ forms a submanifold $\sptqr$, equipped with the unique metric compatible torsion-free connection $\bm{\twod}$ satisfying $\twod_A q_{BC} = 0$. To completely characterize the extrinsic geometry of the null surface, we need the triplet $(\theta_{ab}, \Omega_a, \k)$, where $\theta_{ab}$ is the second fundamental form of $\hech$ defined as \cite{Gourgoulhon:2005ng},
\begin{eqnarray}
	\theta_{ab} = \fr q_{a}^{~i} q_{b}^{~j} \pounds_{\bm{l}}q_{ij} = q_{a}^{~i} q_{b}^{~j} \nabla_i l_j ~.
\end{eqnarray}
Performing an irreducible decomposition of the second fundamental form gives us two new extrinsic quantities derived from its trace and trace-free part.
\begin{align}
	\theta_{ab} = \fr q_{ab} ~\th + \sigma_{ab} ~.
\end{align}
The trace of $\theta_{ab}$ is called the expansion scalar for the null vector field as it correctly quantifies the fractional rate of change of the area element of $\hech$ when evolved along the null generators,
\begin{equation}
	\th = q^{ab} \theta_{ab} =  \frac{1}{\sqrt{q}} \frac{d}{d \ll} \sqrt{q} ~,
	\label{thdl1}
\end{equation}
where $\ll$ represents the non-affine parameter for the outgoing null generators $\bml$. The trace-free shear tensor is then defined as,
\begin{align}
	\sigma_{ab} = q_{a}^{~i} q_{b}^{~j} \nabla_i l_j - \fr q_{ab} \th ~.
\end{align}
The Hajicek one-form or the twist field $\Omega_a$ corresponding to the generators $\bml$ of $\hech$ defined as \cite{Gourgoulhon:2005ng},
\begin{align}
	\Omega_a = -q_a^{~b} (k_c\nabla_b l^c) ~.
	\label{Hajicek}
\end{align}
In principle, the Hajicek one form is the projection of something called the rotation one-form $\omega_a$ onto the submanifold $\sptqr$, defined as \cite{Gourgoulhon:2005ng},
\begin{align}
	\omega_a = -\Pi_a^{~b} k^c \nabla_b l_c = -k^b \nabla_a l_b - l_a (k^b k^c \nabla_b l_c) = l^b \nabla_b k_a ~.
	\label{rotnoneform}
\end{align} 
Finally, $\k$ denotes the non-affinity parameter or surface gravity associated with the null geodesics $\bml$ of $\hech$,
\begin{align}
	l^b \nabla_b\ l^a = \k l^a ~.
	\label{geodeqn}
\end{align} 

To connect the gravitational dynamics on the null surface with fluid dynamical behavior, it will be necessary to consider the evolution laws of the extrinsic quantities defined on $\hech$. In that regard, we consider first the evolution of the second fundamental form along the null generators. It can be shown \cite{Gourgoulhon:2005ng,Poisson:2009pwt},
\begin{align}	
	q_a^{~i} q_{b}^{~j} \Big(\pounds_{\bml} \theta_{ab} \Big) = \k \theta_{ab} + \theta_{ai} \theta^{i}_{~b} - q_a^{~i} q_{b}^{~j} C_{minj}l^m l^{\rm N} - \fr q_{ab} R_{ij} l^i l^j ~,
	\label{thetaevol} 
\end{align}
where $C_{abcd}$ is the Weyl tensor for $\sptmr$. Taking the trace of \eqref{thetaevol} gives us the celebrated null Raychaudhuri equation (NRE),
\begin{align}
	l^b \nabla_b \theta - \kappa \th^2 + \sigma_{ab} \sigma^{ab} + R_{ab}l^a l^b = 0 ~.
	\label{NREeqn}
\end{align}
The evolution of the Hajicek one-form along the null generators is given by the following equation,
\begin{align}
	q_a^{~b}\pounds_{\bml}\Omega_b+\theta~\Omega_a- \twod_a\Big(\frac{\theta}{2}+\kappa\Big)+~^{(2)}D_i\sigma^i_a=R_{mn}l^m q_a^{~n}~.
	\label{DNSeqn}
\end{align}
Upon using the Einstein field equations and trading off the Lie-derivative with a spatial derivative operator $D_t$ acting on the Hajicek one-form defined as,
\begin{align}
    D_t \Omega_a \equiv q_a ^{~b} l^c \nabla_c \Omega_b = q_a^{~b} \pounds_{\bml} \Omega_b - \theta_a^{~b} \Omega_b ~,
\end{align}
and restricting to the spatial coordinates of $\sptqr$, the above \eqref{DNSeqn} gives us the Hajicek equation \cite{PhysRevD.33.915}. 

Before ending this geometrical description of the null surface, it is important to mention that we are obviously considering a null surface at finite distances of the spacetime and not null infinity at the conformal boundary of the spacetime conformal completion.

\section{Searching EM tensor on null hypersurface}\label{secBYformulation}
In our pursuit of finding the quantity that might play the role of EM tensor for null hydrodynamics, let us make some educated guesses. Since we are looking for the (hydro-) dynamics of the null surface, we must look for those quantities that are related to the dynamical equation of the spacetime. But, it cannot be the Einstein tensor itself as the different projections of $G_{ab}$ have been studied rigorously. Also, it is known that for a timelike/spacelike hypersurface, the true dynamical variables are the induced metric tensors and the true dynamical information lies in their conjugate quantities. For a timelike surface, the conjugate quantity is also known as the Brown-York (BY) tensor, which is related to several physical quantities of black holes. Furthermore, it can be shown that the BY tensor is conserved in the absence of any external matter. Hence, the BY tensor seems to be a promising candidate. However, due to the degeneracy of the null hypersurface, one cannot define an induced metric on a null surface. Therefore, the direct route of finding the BY tensor for the null surface does not work. In the following, we discuss the important arguments for identifying the  BY tensor for the usual timelike surface. This will serve the purpose of laying the path to obtain the BY-like tensor on the null surface.

\subsection{BY tensor: laying the path for null counterpart}
The Brown-York formalism of defining surface stress tensor is based on the Hamilton-Jacobi principle of classical mechanics. As has been discussed earlier, the original Brown-York (BY) formalism and, thereby, the BY energy-momentum tensor have been formulated for a non-null hypersurface \cite{Brown:1992br,Brown:2000dz}. The formalism crucially depends on the well-posedness of the action and, therefore, it is not defined solely for the Einstein-Hilbert action without considering suitable boundary term (see \cite{Bhattacharya:2023ycc}, where this formalism has been extended for the scalar-tensor theory). It has been observed \cite{Padmanabhan:2014lwa} that the variation of the Einstein-Hilbert (EH) action along with the Gibbons-Hawking-York (GHY) boundary term yields the following result
\begin{align}
\d \A_{\rm WP} =\d \A_{\rm EH}+\d \A_{\rm GHY}=
\d\Big(\frac{1}{16\pi}\int_{\V}\mg R\ d^4x-\frac{\epsilon}{8\pi}\int_{\p\V}\sqrt{\mathbf{h}} \mathbf{K} d^3x\Big)
\no 
\\
=\frac{1}{16\pi}\int_{\V}\mg G_{ab}\d g^{ab}\ d^4x+\epsilon\int_{\p\V}\sqrt{\mathbf{h}} \Big(\mathbf{D}_a \mathbf{T}^a+\boldsymbol{\pi}_{ab}\d \mathbf{h}^{ab}\Big)d^3x~, \label{deltaawp}
\end{align}
where $\mathbf{h}_{ab}=g_{ab}-\epsilon \mathbf u_a\mathbf u_b$ is the induced metric of an arbitrary timelike/spacelike surface (with normal being $u_a$) which encloses the boundary, $\mathbf{h}$ is the determinant of the induced metric, $\epsilon=\mathbf{u}^i\mathbf{u}_i =\pm 1$ depending upon the surface (spacelike/timelike). Also, in the above \ref{deltaawp}, the well-posed action ($\A_{\rm WP}$) is shown to have consisted of the Einstein-Hilbert action ($\A_{\rm EH}$) along with the GHY boundary term. Furthermore, we use bold text for the quantities of the general surface (timelike or spacelike) to distinguish them from the quantities of the timelike surface which is discussed below. In the above \ref{deltaawp}, $\boldsymbol{\pi}_{ab}$ can be identified as the gravitational momentum, which is given as
\begin{align}
\boldsymbol{\pi}_{ab}=\frac{1}{16\pi}\Big(\mathbf{K} \mathbf{h}_{ab}-\mathbf{K}_{ab}\Big)~,\label{piab}
\end{align}
where $\mathbf{K}_{ab}=-\mathbf{h}^i_a\mathbf{h}^j_b\na_i\mathbf u_j=-\mathbf{h}^i_a\na_i\mathbf u_b$ is the extrinsic curvature of the surface and $\mathbf{K}$ is its trace. The total three-derivative term of \ref{deltaawp},  $\mathbf{D}_a\mathbf{T}^a$ (where $\mathbf{T}^a=\mathbf{h}^a_i\mathbf u_j\d g^{ij}/16\pi$, and the definition of the three-derivative on the surface is given as $\mathbf{D}_aA^b=\mathbf{h}_a^i\mathbf{h}^b_j\na_i(\mathbf{h}^j_kA^k)$, where $A^b$ is an arbitrary vector) can be neglected on the surface and when one extremizes the action.

On a timelike boundary $\mathcal{T}$ (characterised by the spacelike normal $s^a$, thereby $\epsilon=+1$, and the induced metric $h_{ab}=g_{ab}-s_as_b$), the onshell variation of the well-posed Lagrangian ($\L_{\rm WP}$, where $\A_{\rm WP}=\int_{\V}\mg L_{\rm WP}~d^4x$) with respect to the induced metric is given as 
\begin{equation}
\frac{\d(\mg L_{\rm WP})}{\d h^{ab}}\Big|_{\mathcal{T}}=\pi_{ab}\Big|_{\mathcal{T}}=-\frac{T_{ab}^{\rm (BY)}}{2}~.
\end{equation}
Thus, $T_{ab}^{\rm (BY)}$ resembles to the expression of the energy-momentum tensor of the external matter field \textit{i.e.}
\begin{align}
T_{ab}^{\rm (BY)}=-2\frac{\d(\mg L_{\rm WP})}{\d h^{ab}}\Big|_{\mathcal{T}}=\frac{1}{8\pi}\Big(K_{ab}-Kh_{ab}\Big)~. \label{BYTLIKE}
\end{align}
Hence, $T_{ab}^{\rm (BY)}$ is known as the surface energy-momentum tensor A.K.A. Brown-York (energy-momentum) tensor. It can be shown that the three-derivative of the BY tensor is given as 
\begin{align}
D_aT^{ab}_{\rm (BY)}=-T^{ac}_{\rm (ext)}s_ah^b_c~,
\end{align}
where $T^{ab}_{\rm (ext)}$ corresponds to the energy-momentum tensor of the external matter source. In the absence of the external matter, one can obtain $D_aT^{ab}_{\rm (BY)}=0$, which resembles to the conservation of the external energy-momentum tensor $\nabla_aT^{ab}_{\rm (ext)}=0$~.
It can be shown that the Brown-York tensor, as defined in \ref{BYTLIKE}, is related to several quasi-local parameters of the black holes such as quasi-local energy (popularly known as the Brown-York energy), quasi-local mass (Brown-York mass), angular momentum density etc. (for details see \cite{Bose:1998yp,Cote:2019fkf}).

\subsection{BY-like tensor on a null surface}
The above formulation of BY tensor and thereby defining the quasilocal BY parameters has been previously formulated in \cite{Brown:1992br,Brown:2000dz}, which is valid for the timelike surfaces. Since our focus is on the null surface, we look for the analogous term of the BY tensor for the null surface. However, as discussed earlier, obtaining a BY-like tensor in a direct manner (i.e. conjugate quantity of the induced metric) is not possible for the null surface as one cannot define an induced metric in this case. In addition, the suitable boundary term for the null surface was not known until the recent works \cite{Parattu:2015gga,Parattu:2016trq,Lehner:2016vdi,Oliveri:2019gvm,Aghapour:2018icu,Chandrasekaran:2020wwn}. In the following, we define a BY-like tensor in the following, which has been discussed in the recent papers \cite{Jafari:2019bpw,Chandrasekaran:2021hxc,Bhattacharya:2023ycc}.

Following \cite{Parattu:2015gga}, one can define suitable boundary term for the null surface and the variation of the total action can be obtained as 
\begin{align}
\d\A_{\rm WP}^{\rm (null)}=\d\A_{\rm EH}+\d\A_{\rm boundary}^{\rm (null)}=\frac{1}{16\pi}\d\int\mg R~d^4x+\frac{1}{8\pi}\d\int_{\H}\mg\Big(\theta+\k\Big)d^3x~,
\no 
\\
=\frac{1}{16\pi}\int\mg G_{ab}\d g^{ab}\ d^4x+\int_{\H}\Big(\frac{1}{16\pi}\p_a(\mg\Pi^a_{~b}l^b_{\bot})+\mg(\P_{ab}\d q^{ab}+\P^{(l)}_a\d l^a)\Big)d^3x~, \label{nullvariation}
\end{align}
where $l^a_{\bot}=\d l^a+g^{ab}\d l_b$. In addition, $\P_{ab}$ and $\P^{(l)}_a$ are the conjugate quantities of $q^{ab}$ and $l^a$ respectively, which is given as
\begin{align}
\P_{ab}=\frac{1}{16\pi}\Big(\theta_{ab}-(\theta+\kappa)q_{ab}\Big)~,~~~~~~~~~~~\P^{(l)}_a=\frac{1}{8\pi}\Big((\theta+\kappa)k_a+\omega_a\Big)~,
\label{B1}
\end{align}
Unlike the timelike hypersurface, where the true dynamical variable is only the induced metric $h^{ab}$, in this case, the dynamical degrees of freedom are both $q^{ab}$ and $l^a$ \footnote{In four spacetime dimensions, $g_{ab}$ has ten independent components. For timelike surface, the induced metric has six independent components, which can be regarded as the true dynamical variables. On the other hand, for a null surface $\H$, $q_{ab}$ has three independent components as well as $l^a$ has three independent components, making total dynamical degrees of freedom again as six. For details, see \cite{Parattu:2015gga}}.

Interestingly if one investigates the projection $q^a_c G_{ab}l^b$, which upon use of Einstein's equation leads to famous DNS equation \ref{DNSeqn},  includes the information of above $\P_{ab}$ and $\P^{(l)}_a$. Indeed the DNS equation can be expressed in terms of these conjugate momenta. Following the details of Appendix \ref{DNS_PP}, one can show that 
\begin{eqnarray}
&&q^c_a\pounds_l\Big(\sqrt{q} \P^{(l)}_c\Big) = \frac{\sqrt{q}}{8\pi} \Big(\Omega_a\theta + q^c_a\pounds_l \Omega_c\Big)~;
\nonumber
\\
&&\P_{ab} = \frac{1}{16\pi}\Big[\sigma_{ab} - \Big(\frac{1}{2}\theta+\kappa\Big) q_{ab}\Big]~.
\label{DNS_1}
\end{eqnarray}
Use of these in \ref{DNSeqn} one finds a new form of DNS equation as
\begin{equation}
q^c_a\pounds_l\Big(\sqrt{q} \P^{(l)}_c\Big) + 2\sqrt{q} ^{(2)}D_c\P^c_a = \sqrt{q}G_{cd}l^c q^d_a~.
\label{DNS_P}
\end{equation}
This clearly indicates that the required energy-momentum tensor which will lead to the above should be constructed out of these conjugate variables.

 Therefore, the quantity which we are looking after (i.e. BY-like tensor for the null surface), must contain the conjugate quantities of both $q_{ab}$ and $l^a$. In literature, such quantity has been identified in the following form \cite{Chandrasekaran:2021hxc}:
\begin{align}
T^a_{~~b}|_{\rm (null)}=2q^{ai}\P_{ib}+l^a\P^{(l)}_b=\frac{1}{8\pi}\Big(W^a_{~~b}-\Pi^a_{~b} W\Big)~, \label{BYnull}
\end{align}
where $W^a_{~~b}=\theta^a_b+l^a\omega_b$ and $W=\theta+\kappa$~. The above expression of BY-like tensor in \ref{BYnull} resembles to the BY tensor of the timelike surface where $K^a_b$ is replaced by $W^a_{~~b}$ and $h^a_b$ is replaced by $\Pi^a_{~b}$~. In the following, we show that this is the same quantity, which we are looking after.

\section{Hydrodynamic description on a null surface: A formulation based on EM tensor} \label{secGRAVFLUID}

Let us now establish a one-to-one correspondence between the fluid energy-momentum tensor and the null Brown-York tensor which has been defined in the previous section. At first, taking a cue from fluid dynamics (see the text below \ref{confluid}), we identify the energy, momentum and spatial stress from different components of the null BY-like tensor, which can be obtained as 
\begin{eqnarray}
&&\epsilon_{\rm null}=-T^a_{~~b}|_{\rm (null)}k_al^b=-\frac{\theta}{8\pi}~,
\no 
\\
&&p^c=-T^a_{~~b}|_{\rm (null)}k_aq^{bc}=-\frac{\Omega^c}{8\pi}~,
\no 
\\
&&s^{ab}=q^a_cq^{bd}T^c_{~~d}|_{\rm (null)}=2\eta\sigma^{ab}+q^{ab}(\zeta\theta-P)~, \label{BYTPROJ}
\end{eqnarray}
where $\Omega^a=q^{ab}\omega_b$, $\eta=1/16\pi$, $\zeta=-1/16\pi$, $\sigma^{ab}=\theta^{ab}-q^{ab}\theta/2$, $\theta=\theta$ and $P=\kappa/8\pi$. In what follows, we shall check the consistency of such identifications. Firstly, note that one can show (for details, see \ref{CONNULL})\footnote{Here $\D^a$ represents the differentiation operator that is compatible with the projection tensor $\Pi^a_{~b}$. $\D_aA^b=\Pi^i_{~a}\Pi^b_{~j}\na_i(\Pi^j_{~k}A^k)$ and $\D_aB_b=\Pi^i_{~a}\Pi^j_{~b}\na_i(\Pi^k_{~j}B_k)$. Hence, it can be proved that $\D_a\Pi^a_{~b}=0$, implying that $\D_a$ is compatible with the projection tensor $\Pi^a_{~b}$}
\begin{align}
\D_aT^a_{~~c}|_{\rm (null)}=\frac{1}{8\pi}R_{ab}\Pi^a_{~c}l^b~, \label{connullBY}
\end{align}
 Thus, in the absence of the external matter field, upon use of Einstein's equation one obtains $\D_aT^a_{~~c}|_{\rm (null)}=0$, which is analogous to the conservation of fluid energy-momentum tensor as described in \ref{confluid}. Note that such is conserved upon use of Einstein's equation of motion and therefore the conservation of EM tensor carries the dynamics of gravity. In addition, from the above relation \ref{connullBY}, we can obtain the following results for the gravitational fluid dynamics. Firstly, contracting \ref{connullBY} with the auxiliary null vector one finds
\begin{align}
k^c\D_aT^a_{~~c}|_{\rm (null)}=0~. \label{connullBY1}
\end{align}
The above equation indicates that the projection of \ref{connullBY} along the auxiliary null vector vanishes even without using the dynamical information of the null surface (i.e. without using the equation of motion). Hence, it can be considered as a mere geometrical identity of a null surface. Since it does not carry any dynamical information, we do not expect to have any significance from the viewpoint of null hydrodynamics, as we expect the latter to be related to the dynamics of the null hypersurface.  On the contrary, 
contracting \ref{connullBY} with the induced metric of $\B^{\rm (null)}$ provides
\begin{align}
q^c_d\D_aT^a_{~~c}|_{\rm (null)}=\frac{1}{8\pi}R_{ab}q^a_dl^b~.
\label{connullBY2}
\end{align}
Use of vacuum Einstein's equation leads to vanishing of the right hand side and then it must be analogous to \ref{momconfluid}. Therefore we expect that the above one must give dynamics of fluid-description of gravity.
Upon expanding the left-hand side of \ref{connullBY2} and replacing $R_{ab}$ with the energy-momentum tensor corresponding to the external matter field (\textit{i.e.}$G_{ab}=8\pi G T_{ab}^{(ext)}$), we obtain the DNS equation \ref{DNSeqn}. The important point to note is that we have obtained the DNS equation (i.e. \ref{DNSeqn}) conventionally by taking the appropriate projection of the conservation of energy-momentum tensor. In other words, \ref{DNSeqn} is analogous to \ref{momconfluid}. In addition, the fluid parameters are also identified in a conventional manner, i.e. from the components of the null BY tensor.
For a long time, \ref{DNSeqn} has been correlated with the momentum-conservation relation of the Navier-Stokes equation, i.e. \ref{MOMCONFIN}, where we find that our earlier identifications in \ref{BYTPROJ} become consistent. With our previous identification, in the absence of external matter,  \ref{DNSeqn} looks as
\begin{align}
q^{\rm N}_a\pounds_l p_n-\theta p_n-\zeta~^{(2)}D_a\theta+~^{(2)}D_aP-2\eta~^{(2)}D_i\sigma^i_a=0~, \label{dnsnull2}
\end{align}
where, $P$ can be identified as $P=\kappa/8\pi$. Also, the shear viscosity coefficient ($\eta$) and the bulk viscosity coefficient ($\zeta$) can be identified as $\eta=1/16\pi$ and $\zeta=-1/16\pi$. The above equation, famously known as the Damour-Navier-Stokes equation, resembles to the momentum conservation equation of fluid dynamics \cite{Gourgoulhon:2005ng}. The only difference, however, is the presence of the Lie-derivative of the momentum density instead of the convective derivative. In the following section, when the above equation \ref{dnsnull2} is re-written in terms of the adapted coordinate, it can be shown that the same equation (with some conditions) provides a direct resemblance with the momentum conservation equation of a non-relativistic fluid. Using the identifications of the fluid parameters, given in \ref{BYTPROJ}, one can find the exact values of the same for the null-null structure of a spacetime. To illustrate this in \ref{AppC} we calculate the fluid parameters corresponding a null surface defined on Schwarzschild, Kerr and FLRW spacetimes, respectively.

Now the question arises whether there exists an analogous relation to the energy–conservation equation for null hydrodynamics. To investigate this, we take another projection of \ref{connullBY}, contracting it with the null generator $l^c$, which gives
\begin{align}
l^c\D_aT^a_{~~c}|_{\rm (null)}=\frac{1}{8\pi}R_{ab}l^al^b~. 
\label{connullBY3}
\end{align}
Expanding the left-hand side yields the Raychaudhuri equation \ref{NREeqn}, a classical geometrical identity in general relativity. This equation has long been employed in the proof of the black-hole area theorem and hence occupies a central role in the thermodynamic interpretation of horizons. Here, however, we show that it also admits a fluid interpretation. In vacuum the right-hand side of \eqref{connullBY3} vanishes, so the equation becomes directly analogous to \eqref{enconfluid}. Therefore we expect it to be interpretable as an energy–conservation relation within the fluid description of gravity. This correspondence is particularly transparent in adapted coordinates; in the next section we will demonstrate that \ref{connullBY3} indeed takes the form of a fluid energy–conservation equation. Thus, in a fully covariant manner, we connect the gravitational dynamics on $\hech$ with fluid dynamics beginning from a fluid energy–momentum tensor in the null-foliation setting.

Before proceeding, we remark on the distinction between our approach and the recent literature on Carrollian fluids. It should be pointed out that the previous works, which interprets in terms of Carroll fluid, have been done under a different geometrical construction. That particular construction is essentially that of a stretched horizon, employed both in the membrane and the Carroll paradigms, where the spacetime about the internal null boundary $\hech$ is foliated by a stack of timelike surfaces. Now, under this stretched horizon purview, the  analysis leads to a fluid description of the null surface \cite{thorne1986black, PhysRevD.33.915,Parikh:1997ma,Freidel:2022vjq}. As can be evident, these constructions rely on the existence of a family of timelike stretched surfaces/horizons, with the null hypersurface being a limiting case in this family of stretched horizons. That is, geometrically the neighborhood of the null surface is foliated by a family of timelike hypersurfaces (as opposed to null surfaces as in our case). 

In the membrane paradigm, the stretched horizons provide a quasi-local description for the physics of the black hole event horizon with respect to timelike observers arbitrarily close to the actual event horizon. These observers then attribute these stretched membranes to a viscous fluid description. Suitable projection of the Einstein field equation on the stretched horizons can be interpreted as fluid dynamic laws with the stretched horizon being attributed to fluid parameters such as pressure, energy density, heat flux, and appropriate transport coefficients \cite{thorne1986black, Parikh:1998mg}. Hence in the membrane paradigm, timelike and null surfaces have been treated on an equal footing. The true physical degrees of freedom encoded on the null boundary can only be accessed by considering small deviations from it. These gravitational degrees of freedom can be accessed by such arbitrarily close timelike observers on the stretched membranes. It has been shown that the radial ($1/r$) expansion around the asymptotic null infinity encodes higher-spin symmetries and conservation laws of the null infinity \cite{Freidel:2021qpz, Freidel:2021dfs, Freidel:2021ytz}. One fundamental issue regarding the membrane paradigm is that while taking the null limit of this sequence of stretched horizons, there occur divergences due to infinite redshift effects. The induced metric and the connection on the stretched horizon become singular when the limiting process to the null boundary is taken. In the context of stretched horizons, this puzzle was resolved by demonstrating that the near horizon geometry and the Einstein field equation on the horizon can be understood in terms of Carrollian geometry \cite{Ciambelli:2018xat, Ciambelli:2018ojf, Petkou:2022bmz, Freidel:2022bai}. Subsequent work in this gave rise to the Carrollian membrane paradigm.

In the Carrollian membrane paradigm, the vicinity of the null boundary $\hech$ (located at a finite distance) is foliated by a family of three-dimensional timelike surfaces $\mathcal{T}$. Such a family of timelike surfaces is defined by the scalar function $r(x^a) = \text{constant} > 0$. The null boundary is obtained in the limit $r(x^a) = 0$. That is, essentially, the null limit to $\hech$ is obtained by taking the limit $r \rightarrow 0$. However, such a family $\mathcal{T}$ is not foliated by spacelike hypersurfaces $\Sigma_t$ (which would in turn provide the notion of a time evolution vector field $\bm{t}$ and hence predispose the structures $\hech$ and $\mathcal{T}$ with a Galilean picture). Instead, conformal Carroll structures are imposed on these timelike surfaces $\mathcal{T}$ as well as the null boundary $\hech$ \cite{Freidel:2022vjq, Freidel:2024emv}. The normal form to these Carroll horizons $\mathcal{T}$ is given by 
\begin{eqnarray}
	\underline{\bm{s}} = e^{\bar{\alpha}} \underline{\bm{d}}r ~,
\end{eqnarray}
with $\bar{\alpha}$ being a smooth scalar function on the spacetime manifold.
Such a Carroll geometry is implemented via the null rigged construction \cite{Mars:1993mj, Mars:2013qaa}, by using a null rigging vector $\bm{k}$ which is dual to the normal form and hence transverse to $\mathcal{T}$. The null rigged structure, by choosing the null vector $\bm{k}$, is regular for both timelike and null surfaces. This removes the issue of singularities encountered in the membrane paradigm while taking the null limit \cite{Freidel:2022vjq}. The normal one form $\underline{\bm{s}}$ is obviously non-null given by,
\begin{align}
	\bm{g}^{-1}({\underline{\bm{s}},\underline{\bm{s}}}) \equiv 2 \t\rho ~,
\end{align}
with $\t\rho$ being a smooth scalar in $\sptmr$. The Carroll structure becomes null only on the null boundary obtained by taking $\t\rho =0$. In that case, the normal vector $\bm{s}$ coincides with the null generators $\bm{l}$ of $\hech$. With such a Carrollian structure implemented on $\mathcal{T}$, an appropriate Carrollian fluid energy-momentum tensor can be described on the stretched horizon \cite{Freidel:2022vjq, Freidel:2024emv}.

At this point, it becomes quite imperative to discuss the differences in the approach of stretched membranes and those employed in our present study. At the very outset, the obvious constructional difference is regarding the foliation of $\sptmr$ about the null boundary $\hech$. The stretched surface paradigm involves foliating the spacetime in the vicinity of $\hech$ by timelike hypersurfaces. In our work, following Carter, we impose a null foliation of spacetime. By this, we avoid the issues of singularity that arise in the membrane paradigm picture by taking the null limit. However, as we show in the next section, in line with the membrane paradigm \cite{thorne1986black, PhysRevD.33.915,Parikh:1997ma}, we can employ a $3+1$ decomposition of the null boundary $\hech$. This gives rise to a preferred time evolution vector field $\bm{t}$ take breaks general covariance bringing into the structure of $\hech$, the Galilean picture. This is avoided in the Carrollian paradigm. Another minor technical difference that arises in the constructions is that of the nature of the null rigging vector $\bm{k}$. For stretched horizon method, the null rigging vector is assumed to generate null geodesics i.e. they satisfy the auto-parallel equation ($\nabla_{\bm{k}} \bm{k} = \t{\k} \bm{k}$, $\t{\k}$ being the non-affinity parameter). This is also the same case as employed in the construction of Gaussian null coordinates w.r.t. the null boundary $\hech$. However, in our case of the null foliation of spacetime, the auxiliary null vector field is not constrained to be geodesic.

\section{Is there any non-relativistic counterpart?}\label{Aside}
In a non-relativistic scenario and in the absence of external forces, the continuity equation, energy conservation relation, and momentum conservation relations correspond to NS fluid. They are given as follows \cite{HegadeKR:2023glb}:
\begin{eqnarray}
&&D^{\rm N}\rho+\rho\theta^{\rm N}=0~,
\no 
\\
&& \rho D^{\rm N}\epsilon^{\rm N}+ P\theta^{\rm N}+\p_{\mu}Q^{\mu}-\zeta(\theta^{\rm N})^2-2\eta(\sigma^{\rm N})^{\mu \nu}\sigma^{\rm N}_{\mu \nu}=0~,
\no 
\\
&&\p^{\mu}P+\rho D^{\rm N}v^{\mu}-\p_{\nu}\Big[\zeta\theta^{\rm N}\delta^{\mu \nu}+2\eta(\sigma^{\rm N})^{\mu \nu}\Big]=0~, \label{allfluideqn}
\end{eqnarray}
where
\begin{eqnarray}
&&D^{\rm N}\equiv\p_t+v^{\mu}\p_{\mu}~,
\no 
\\
&&\sigma^{\rm N}_{\mu \nu}\equiv\p_{(\mu}v_{\nu)}-\frac{1}{3}\delta_{\mu \nu}\theta^{\rm N}~,
\no 
\\
&&\theta^{\rm N}=\p_{\mu}v^{\mu}~.
\end{eqnarray}
Here the index ``N" is used for Navier-Stokes (non-relativistic) fluid, which is used to distinguish the physical quantities from the gravitational counterpart. Furthermore, $\rho$ is the mass density of the fluid (mass per unit volume), $v^{\mu}$ represents the 3-velocity components of the fluid ($\mu=1,2,3$), $P$ is the thermodynamic pressure, $\epsilon^{\rm N}$ represents specific internal energy (per unit mass) in the non-relativistic limit, $Q^{\mu}$ is the heat flux vector, $\zeta$ is the bulk viscosity coefficient, $\eta$ is the shear viscosity coefficient, $\theta^{\rm N}$ represents the expansion scalar, $\sigma^{\rm N}_{\mu\nu}$ represents the shear viscosity tensor, and $D^{\rm N}$ corresponds the convective derivative. If one uses the continuity equation in the energy conservation and momentum conservation relation, one obtains the modified form of energy and momentum conservation relations, where the energy conservation relation is given as
\begin{align}
&&  D^{\rm N}(\rho\epsilon^{\rm N})+\Big(\rho\epsilon^{\rm N}+ P\Big)\theta^{\rm N}+\p_{\mu}Q^{\mu}-\zeta(\theta^{\rm N})^2-2\eta(\sigma^{\rm N})^{\mu \nu}\sigma^{\rm N}_{\mu \nu}=0~,
\label{ENCONFIN}
\end{align}
and the momentum conservation relation is given as
\begin{align}
&& D^{\rm N}\Big(\rho v^{\mu}\Big)+\rho v^{\mu}\theta^{\rm N}+\p^{\mu}P-\p_{\nu}\Big[\zeta\theta^{\rm N}\delta^{\mu \nu}+2\eta(\sigma^{\rm N})^{\mu \nu}\Big]=0~. 
\label{MOMCONFIN}
\end{align}
\ref{MOMCONFIN} provides the dynamics of the fluid while \ref{ENCONFIN} describes its kinematics. In what follows, we will show that the analogous relations of \ref{ENCONFIN} and \ref{MOMCONFIN} exist in gravitational dynamics on a generic null hypersurface, provided we pick out an adapted coordinate system w.r.t. $\hech$. More importantly, those relations are obtained in a conventional manner i.e. taking different projections of conservation of suitably chosen gravitational energy-momentum tensor, similar as described in \ref{enconfluid} and \ref{momconfluid} for usual fluid. 

We now discuss the notion of adapting a coordinate system with respect to our integrable null surface $\hech$ as detailed out in \cite{Gourgoulhon:2005ng}. As mentioned before, owing to the $3+1$ foliation of $\hech$ by the family of spacelike surfaces $\Sigma_t$ allows us to coordinatize spacetime manifold in the vicinity of $\hech$ by $(t, x^1,x^2,x^3)$. Adapting the coordinates to the null surface means that we can fix the location of $\hech$ by the choice $x^{1}= 0$. Along the time evolution vector field $\bm{t} = \bm{\p_t}$, the transverse/angular coordinates of $\bnl$ are fixed.
It can be shown \cite{Gourgoulhon:2005ng}, that on the null surface we have, $\bm{l} \overset{\hech}{=} \bm{t} + \bm{V}$, where $\bm{V}$ is spatial vector field that is tangent to $\bnl$. With respect to the adapted coordinate system, hence, we have $l^{\t \mu} \overset{\hech}{=} (1, V^A)$. The line element on the null surface in this adapted coordinate system can be expressed as \cite{Gourgoulhon:2005ng, Padmanabhan:2010rp},
\begin{align}
	ds^2_{|\hech} = q_{\t \mu \t \nu} dx^{\t \mu} dx^{\t \nu} = q_{AB}(dx^A - V^A dt)(dx^B - V^B dt) ~.
	\label{indhech}
\end{align}
It can be shown \cite{Gourgoulhon:2005ng} that w.r.t. the adapted coordinate system introduced above, the second fundamental form takes the structure,
\begin{align}
	\theta_{AB} \overset{(\mathcal{M}, \bm{g}, \bm{\nabla})}{=} \fr( \p_t q_{AB} + \twod_A V_B + \twod_B V_A) ~.
	\label{manip34}
\end{align} 
Now, it could be possible that for a specific choice of the adapted coordinates, we could have the constraint that $\p_t q_{AB} = 0$ i.e. the induced metric $q_{AB}$ is stationary or independent of the time evolution parameter $t$. Then the second fundamental form of the null surface $\hech$ takes exactly the same structure as that of the stress tensor of a two-dimensional viscous fluid with velocity $V^A$. This part will be used to relate the gravitational equations with NS fluid equations \ref{ENCONFIN} and \ref{MOMCONFIN}.

For the case of Einstein gravity, upon analyzing the evolution equation for the Hajicek one-form \ref{DNSeqn} in the adapted coordinate system w.r.t. $\hech$ and using that $\pounds_{\bml} \Omega_a\overset{\hech}{=} \pounds_{\bm{t}} \Omega_a+ \pounds_{\bm{V}} \Omega_a$, we have,
\begin{align}
    \p_t \Omega_a + V^b \twod_b \Omega_a + \Omega_b \twod_a V^b + \th \Omega_a - \twod_a\Big(\frac{\theta}{2}+\kappa\Big)+~^{(2)}D_i\sigma^i_a = 8 \pi q_a^{~i} \mt_{ij}l^j ~.
    \label{dns}
\end{align}
This is the well-known Damour-Navier-Stokes equation \cite{Damour:1979wya, damour1982surface}. In fact, it was shown by Padmanabhan \cite{Padmanabhan:2010rp}, that analyzing the DNS equation \ref{dns} with respect to a boosted local inertial frame, the extra term $\Omega_b \twod_a V^b$ drops out and it takes the structure of a two dimensional viscous Navier-Stokes fluid, identical to \ref{MOMCONFIN} for a specific choice of fluid parameters in terms of gravitational quantities.
So following that idea, \ref{dns} becomes (in the absence of external matter)
\begin{align}
    \Big(\p_t + V^B \p_B\Big) \Big(\frac{-\Omega_A}{8\pi} \Big)-\frac{1}{8\pi}\p_B\sigma^B_A+\frac{1}{16\pi}\p_A\theta+\p_A\Big(\frac{\kappa}{8\pi}\Big)=0~,
\end{align}
which, upon using our identification \ref{BYTPROJ} (along with $\eta=1/16\pi$, $\zeta=-1/16\pi$, $\theta=\theta$, and $P=\kappa/8\pi$) takes the form
\begin{align}
    \Big(\p_t + V^B \p_B\Big)p_A-2\eta~ \p_B\sigma^B_A-\zeta\p_A\theta+\p_AP=0~. \label{DNSnoninertial}
\end{align}
In the above \ref{DNSnoninertial}, the discrepancy between Lie-derivative and convective derivative is resolved and \ref{DNSnoninertial} looks exactly like \ref{MOMCONFIN}. Again, let us emphasize the fact that here we have obtained the known DNS equation, which is analogous to the momentum conservation relation, in a conventional manner i.e. by taking the appropriate projection of the conservation of energy-momentum tensor. Moreover, we show that the fluid parameters can be naturally identified from the different components of the energy-momentum tensor.

 
As promised in the previous section, we now demonstrate that \ref{connullBY3}, the null Raychaudhuri equation, obtained as a particular projection of the conservation of the null Brown-York tensor, admits a fluid–dynamic interpretation when expressed in adapted coordinates.
In absence of matter, identifying $l^a=(1,0,\mathbf{V})$, and expanding the left hand side of \ref{connullBY3} one obtains (for details, see \ref{appenencon}),
\begin{align}
\frac{\p}{\p t}\Big(-\frac{\theta}{8\pi}\Big)+V^A~^{(2)}D_A\Big(-\frac{\theta}{8\pi}\Big)+\Big(\frac{\kappa-\theta}{8\pi}\Big)~^{(2)}D_AV^A=\frac{1}{8\pi}\sigma_{ab}\sigma^{ab}-\frac{1}{16\pi}\Big(~^{(2)}D_AV^A\Big)^2~,
\label{energyeqn}
\end{align}
under the restriction $\p_t q_{AB} = 0$.
Now with our previous identifications, which can be written as
\begin{align}
\Big(\frac{\p}{\p t}+V^A~^{(2)}D_A\Big)\epsilon_{\rm null}+\Big(P+\epsilon_{\rm null}\Big)~^{(2)}D_AV^A=2\eta\sigma_{ab}\sigma^{ab}+\zeta\Big(~^{(2)}D_AV^A\Big)^2~.
\label{energyeqnfluid}
\end{align}
This exactly resembles to \ref{ENCONFIN} with the vanishing heat flux. Thus, we not only obtain the energy conservation relation for null hydrodynamics (which was missing earlier in the literature), we obtain it conventionally. This also suggests that our identification of null BY-like tensor as the analogous of the EMT is appropriate and the entire fluid-gravity description can be extracted from the Brown-York-like energy-momentum tensor of the null surface. One only has to take the conventional projection of \ref{connullBY} to obtain the desired relations. Thus this provides a conventional description of the fluid interpretation of gravity. As mentioned earlier, the choice of $\p_t q_{AB} = 0$ implies that in an adapted coordinate system concerning $\hech$, the induced metric on $\bnl$ becomes stationary. It is precisely due to this constraint as shown in \eqref{manip34} that the second fundamental form becomes equivalent to the stress tensor of a viscous two-dimensional null fluid. Moreover, note that the obtained energy- and momentum conservation relations on the null surface, i.e. \ref{energyeqnfluid} and \ref{dnsnull2} (or \ref{DNSnoninertial} in a local inertial frame) matches with \ref{ENCONFIN} and \ref{MOMCONFIN} of the fluid equations. Those fluid equations (i.e. \ref{ENCONFIN} and \ref{MOMCONFIN}) are originally derived from \ref{allfluideqn} by plugging the continuity equation with the energy- and momentum conservation relation. This implies that continuity relation in null hydrodynamics is embedded in \ref{energyeqnfluid} and \ref{dnsnull2} (or \ref{DNSnoninertial}), which we obtain in the following indirect manner.

 The above discussion explores the fact that the gravitational dynamics, reflected through the conservation of BY-like tensor on the null surface, can be interpreted as NS fluid under a few restrictions on the gravity side. With this setup, gravity can be interpreted as a non-relativistic viscous fluid, and the fluid parameters can be determined in terms of gravitational parameters. We already mentioned a few of them. We will see if more such parameters can be further identified. For example, comparison of \ref{energyeqn} with \ref{ENCONFIN} leads us to the identification of the mass-energy density of the effective fluid, which is given by
\begin{align}
	\en = -\frac{\th}{8 \pi}~,
	\label{enrgyden}
\end{align}
where $\rho$ is the analogue of mass density (in this case it is mass per unit area of $\B^{\rm (null)}$). The momentum density is $\rho v_A = -\Omega_A/8 \pi$. Transacting this relation with the fluid velocity $V^A$, we obtain,
\begin{align}
	\rho = - \frac{\bm{\Omega}\cdot \bm{V}}{8 \pi \bm{V}^2} ~.
	\label{dnsty}
\end{align}
Thus, the continuity equation in null hydrodynamics can be framed in the form of the first equation in \ref{allfluideqn} as
\begin{align}
   \Big(\frac{\p}{\p t}+V^A~^{(2)}D_A\Big)\Big(- \frac{\bm{\Omega}\cdot \bm{V}}{8 \pi \bm{V}^2}\Big)- \frac{\bm{\Omega}\cdot \bm{V}}{8 \pi \bm{V}^2} ~^{(2)}D_AV^A=0~.
\end{align}
Comparing \ref{dnsty} with \ref{enrgyden}, we obtain the non-relativistic version of the internal energy density of the fluid being,
\begin{align}
	\epsilon^{\rm N} = -\frac{\th}{8 \pi \rho} =  \frac{\th \bm{V}^2}{\bm{\Omega}\cdot \bm{V}}~.
	\label{eden}
\end{align}
At this point, we employ the following Euler relation among the fluid thermodynamic parameters \cite{callen1991thermodynamics, Gourg},
\begin{align}
	T \bar{s} = \epsilon^{\rm N} + \frac{P}{\rho} - \mu n ~,
	\label{eos}
\end{align}
where $\bar{s}$ is the entropy per unit mass, $ \mu$ is the chemical potential and $n$ is the number density of the fluid. Here, we also make a few identifications. In the emergent paradigm of gravity, the gravitational dynamics of the null surface are observer-dependent \cite{Chakraborty:2015aja, Padmanabhan:2009vy, Padmanabhan:2015lla, Padmanabhan:2007en}. That is concerning an accelerated observer arbitrarily close to the null boundary who perceives a portion of $\hech$ as his/her Rindler horizon, the gravitational dynamics lend themselves such thermodynamic/fluid dynamic connotations. To such an observer, the temperature of the null surface is given by $T = \k/2 \pi$. If we are allowed to identify the chemical potential as being,
\begin{align}
	\mu = \frac{\th \bm{V}^2}{n ~ \bm{\Omega}\cdot \bm{V}} ~,
	\label{chempot}
\end{align}
then using \ref{chempot}, \ref{dnsty} and the fact that $P = \k/8 \pi$ in \ref{eos}, we obtain that the entropy per unit mass of the null fluid is,
\begin{align}
	\bar{s}= -\frac{2 \pi \bm{V}^2}{\ov} = \frac{1}{4 \rho} ~.
	\label{entrden}
\end{align}
Demanding that the entropy per unit mass is positive requires us to impose that $\ov <0 $. Using the fact that $\rho$ is the mass per unit area of the fluid, we obtain that,
\begin{align}
	s = \frac{\bar{s} ~ \rho ~\sqrt{q}}{\sqrt{q}} = \frac{1}{4} ~,
	\label{BekHawk}
\end{align} 
where $\sqrt{q}$ is the area element on the null surface.
This implies that the entropy of the null fluid per unit mass per unit area of the null surface is the usual Bekenstein Hawking entropy density. This shows that the null fluid describing the gravitational dynamics of $\hech$ is consistent with the KSS bound with {$\eta/s = 1/4 \pi$.} The imposition on $\ov < 0 $, further implies via \ref{chempot} and \ref{eden}, that the chemical potential and the energy density of the effective fluid is negative {unless $\theta<0$. However, the fact that the internal energy of gravity is negative is because of its self-attractive nature. This means that the expansion of $\hech$ is positive, which is indeed supplemented by its teleological boundary conditions. The fact that for the null fluid, we have the temperature $T = \k / 2 \pi$ and the pressure $P = \k /8 \pi$, leads us to the equation of state,
\begin{align}
	P = \frac{T}{4} ~.
	\label{eos1}
\end{align}
This equation of state arises for a two-dimensional Bose gas \cite{Skakala:2014eba,Bhattacharya:2015qkt,Cropp:2016ajh} whose chemical potential is negative \cite{Swarup_2004}.  

	One final comment we make is on the nature of a constraint that is employed to bring in the fluid description. In the case of Carrollian hydrodynamics, while considering the surface energy-momentum tensor, the scalar field $\t\rho$ (describing the non-nullness of the normal one form to $\mathcal{T}$) is assumed to be constant over the Carrollian membrane $\mathcal{T}$. This can be suitably done for a choice of the scalar function $\bar{\alpha}$. In our case, we have employed the constraint, that the induced metric on the transverse cut $\bnl$ of $\hech$ is stationary and hence independent of the time parameter. This can be done with a preferred choice of an adapted coordinate system w.r.t $\hech$.

Before going to the discussion section let us conclude this section with the following comments. The membrane paradigm approach, introduced by Price, Thorne, and MacDonald \cite{thorne1986black, PhysRevD.33.915}, was originated from the work of Damour \cite{Damour:1979wya, PhysRevD.18.3598}. Along the same lines as the works done under the membrane paradigm, we also here make a $3+1$ spacetime split thus rendering a preferred choice of a time coordinate. With such a spacetime split and a given choice of adapted coordinates concerning the null surface, we essentially break away from general covariance, while analyzing the field equations on the null surface. There arise certain conceptual issues under such an analysis. The first one is regarding the bulk viscosity of the two-dimensional null fluid being negative, which is unlike ordinary fluids. This means that such a fluid is unstable under perturbations. This is related to the fact that a global null hypersurface tends to expand or contract \cite{damour1982surface}. The NRE, under the adapted coordinate system, is identified with that of the energy conservation equation of a viscous fluid \eqref{energyeqnfluid} with no heat flux. The energy density is proportional to the expansion scalar and the NRE is a nonlinear first-order differential equation for the expansion scalar. Since the energy density is negative due to the self-attractive nature of Einstein's gravity, this implies that the expansion scalar is positive implying that the null surface only expands under evolution along the null generators or time ({under physically motivated energy conditions}). This has no counterpart in ordinary fluid dynamics. To avoid this unbounded growth in the horizon area, teleological boundary conditions have to be imposed \cite{Parikh:1998mg, Bhattacharya:2017dgr}. That is, instead of specifying the initial conditions, one must provide the final conditions (say, for example, the horizon at the end of time-evolution reaches equilibrium defined by vanishing expansion). The growth of the null surface is acausal.


\section{Discussions}\label{secDISCUSS}
The curious relation found by Damour \cite{Damour:1979wya, PhysRevD.18.3598} (i.e. \ref{DNSeqn}) within null foliation formalism tempted physicists to draw the analogy between fluid dynamics with the dynamics of the null surface. However, the DNS equation is not exactly the same as the NS equation as we discussed earlier. Hence, it made physicists skeptical regarding the potential connection between fluid dynamics and the gravitational dynamics on the null surface. For example, Price and Throne do not connect Damour's result with the NS equation \cite{Price:1986yy}. Instead, they refer to it as the Hajicek equation. To resolve the difference between the DNS equation and NS equation, Padmanabhan \cite{Padmanabhan:2010rp} suggested a local inertial frame, in which case the DNS equation and NS equation become identical. However, there are no physical reasons for preferring such a frame. Furthermore, since there are no other relations that show the connection between fluid dynamics and the dynamics of the null surface, it was also not possible to examine whether the identifications of fluid parameters in DNS equations are consistent. In addition, the quantity that might play the role of EMT in null hydrodynamics was also unknown.

To address these issues, we introduce our analysis. Our objective has been to explore and strengthen the study of the emergent paradigm of gravity, specifically from the hydrodynamic setting. We reiterate that previous works have indeed illuminated upon the connections with gravitational dynamics and hydrodynamics on an internal null boundary. Chief among them have been the Membrane paradigm and Carroll hydrodynamics. Both these works insist on studying and analyzing the gravitational dynamics of an internal null hypersurface in terms of appropriate fluid dynamics and their symmetries. Central to both these formalisms have been the idea of studying the gravitational dynamics on the internal null boundary $\hech$ by foliating the spacetime in the vicinity of $\hech$ by a family of timelike hypersurfaces with an appropriate way of taking the null limit to $\hech$. The key difference in this work has been essentially to study the gravitational dynamics of $\hech$ by foliating the neighborhood of $\hech$ by a family of null surfaces. This is mathematically pleasing as there is no need to take a null limit and their are no issues of divergences/singularities that can occur. In tandem with the previous works of the Membrane and Carroll membrane paradigms, we delved ourselves into this venture of filling in the procedures in this set-up where the spacetime in the vicinity of the internal null boundary is foliated by a null family. For this, in such a null foliated construction, we proceeded to use the appropriate null energy momentum tensor. The central theme then is interpreting the conservation of such a well-defined null Brown-York tensor on $\hech$ as providing a hydrodynamical description of the null surface dynamics as that of a two-dimensional null viscous fluid. Knowing that the true dynamical degrees of freedom of the (degenerate) null surface $\hech$ being $\P_{ab}$ and $\P^{(l)}_a$, which are the conjugate variables to $q^{ab}$ and $l^a$, we explored and showed that the gravitational dynamics of $\hech$ can be encoded in terms of these quantities. Using the well-defined null BY tensor defined on $\hech$ which plays the role of the EMT, we have shown how all the central hydrodynamical equations can be derived as (vacuum) conservation laws of such a null BY tensor. For such a null BY tensor, the corresponding gravitational counterparts of energy density, momentum density, and spatial stress tensor can be well defined. If the null surface dynamics is indeed to be interpreted as a null fluid, then these gravitational counter-parts (of energy density, momentum density, and spatial stress) should also be extracted from the fluid dynamical equations as well. This is precisely what we did as consistency checks by interpreting the null surface dynamics as conservation laws of the null BY tensor (in the vacuum case). With the conservation law of the null BY tensor projected onto the spatial cross section $\bnl$, we obtained the DNS equation, which is equivalent to the momentum conservation law in null hydrodynamics. Similarly, projecting the conservation equation of the null BY tensor along the null generators of $\hech$ and exploring the resulting dynamics in a stationary adapted coordinate system, we obtained the energy conservation law of null hydrodynamics. We notice that the energy density of such a null fluid is negative and its heat current vanishes. Our analysis shows that the suitable projections of the conservation law of the null BY tensor give rise to the necessary hydrodynamical laws of the null fluid. For such a null fluid and suitable identifications of temperature and chemical potential, we used the (densitized) Euler relation to indeed verify that the entropy density in the vacuum Einstein case is the usual Bekenstein-Hawking entropy density and that this null fluid satisfies the KSS bound.

 We have hopefully strengthened the picture of emergent gravity by providing a consistent viewpoint of deriving all the necessary hydrodynamical content of the null surface dynamics arising from the conservation law of a well-defined null Brown York tensor. Hence we hope that with this work, the absence of fluid description of gravity on a generic null surface by adopting null foliation in the previous literature, has now been filled.

 A natural question that may arise in this context is what happens when we perform the analysis on null infinity rather than an internal null boundary at finite distances. In this regard, we became aware of a work \cite{Ciambelli:2025mex} dealing precisely with this question, which came after we submitted our manuscript for peer review. In the work, similar to our geometrical construction, the author considered a unified framework by foliating the bulk spacetime with a family of null surfaces and consequently studying the null surfaces within the bulk in the Bondi-Sachs gauge and then sending off this null surface towards future null infinity. Using the Carrollian data on the internal surface, they study the projections of the Einstein's equations (specifically the null Raychaudhuri and the Damour Navier-Stokes ones) on the null surface and show that at null infinity they asymptote to the Einstein's equation for the boundary metric. By assuming that the boundary metric is time independent (which is analogous to our constraint $\p_t q_{AB} = 0$), the author quite interestingly shows that the asymptotic limit of the time component of the conservation of the null Brown-York stress tensor (which effectively yields the null Raychaudhuri equation) leads to the Bondi mass-loss formula \cite{Strominger:2017zoo}. The author provides a hint that asymptotic limit of the Damour Navier-Stokes equation must be related to the angular momentum equation involving the angular momentum aspect \cite{Strominger:2017zoo}. The author also proposes the asymptotic limit of the null Brown-York stress tensor and shows that the phase space of the internal null boundary asymptotes to the Ashtekar-Struebel phase space \cite{Ashtekar:1981bq} of null infinity.

\section*{Acknowledgement}
The research of KB is supported by the New Faculty Seed Grant (NFSG) of BITS Pilani Dubai Campus. The research of SD is supported by the Polish National Science Centre, as part of the OPUS 22 project number 2021/43/B/ST2/02950. The research of BRM is supported by the Science and Engineering Research Board (SERB), Department of Science $\&$ Technology (DST), Government of India, under the scheme Core Research Grant (File no. CRG/2020/000616).

\appendix
\section*{Appendices}

\section{Derivation of Eq. \ref{DNS_1}}\label{DNS_PP}
\renewcommand{\theequation}{A.\arabic{equation}}
\setcounter{equation}{0}
The conjugate momentum $\P_a$, given in \ref{B1}, upon use of $\omega_a=\Omega_a - \kappa k_a$ can be expressed as
\begin{equation}
P_c = \frac{1}{8\pi}\Big(\theta k_c + \Omega_c\Big)~.
\label{A1}
\end{equation}
Then one finds 
\begin{eqnarray}
8\pi q^c_a\pounds_l \P_c &=& q^c_a\pounds_l\Omega_c + \theta q^c_a\pounds_l k_c
\nonumber
\\
&=& q^c_a\pounds_l\Omega_c + \theta q^c_a(\omega_c + k^i\nabla_c l_i)
\nonumber
\\
&=& q^c_a\pounds_l\Omega_c + \theta \Omega_a + \theta k^i\nabla_al_i + \theta l_a k^i k^c\nabla_cl_i - \theta\kappa k_a~.
\label{A2}
\end{eqnarray}
However using an identity $\nabla_ak_i = \Sigma_{ai} - \Omega_a k_i - k_a \omega_i - l_a k^m\nabla_m k_i$ (see Eq. ($5.85$) of \cite{Gourgoulhon:2005ng}), one term of the above, namely $k^i\nabla_al_i$ can be expressed as
\begin{equation}
k^i\nabla_a l_i = - l^i\nabla_ak_i = -\Omega_a+\kappa k_a - l_ak^ik^m\nabla_m l_i~.
\label{A3}
\end{equation} 
Use of this in \ref{A2} yields
\begin{equation}
q^c_a\pounds_l \P_c = q^c_a\pounds_l \Omega_c~.
\label{A4}
\end{equation}
On the other hand using (\ref{B1}) and $\theta = -(1/2) q_{ij}\pounds_l q^{ij}$ we have
\begin{equation}
q^c_a\P_c \pounds_l\sqrt{q} = \sqrt{q} \Omega_a \theta~.
\label{A5}
\end{equation}
Finally, combining \ref{A4} and \ref{A5} we find \ref{DNS_1}.

\section{Conservation of null BY tensor}\label{CONNULL}
\renewcommand{\theequation}{B.\arabic{equation}}
\setcounter{equation}{0}
The expression of null BY tensor is given by \ref{BYnull}.  The differentiation operator $\D_a$ operates on contravariant and covariant vectors as $\D_aA^b=\Pi^i_{~a}\Pi^b_{~j}\na_i(\Pi^j_{~k}A^k)$ and $\D_aB_b=\Pi^i_{~a}\Pi^j_{~b}\na_i(\Pi^k_{~j}B_k)$, which we have mentioned earlier. Therefore, we obtain 
\begin{align}
    \D_a T^a_{~~c}|_{\rm (null)}=\Pi^i_{~a}\Pi^a_{~j}\Pi^k_{~c}\na_iT^j_{~~k}|_{\rm (null)}=\Pi^i_{~j}\Pi^k_{~c}\na_iT^j_{~~k}|_{\rm (null)}~,
\end{align}
which provides straightforwardly as 
\begin{align}
    \D_a T^a_{~~c}|_{\rm (null)}=\Pi^b_{~c}\Big(\na_a T^a_{~~b}|_{\rm (null)}-k^i\na_il_aT^a_{~~b}|_{\rm (null)}\Big)~,
\end{align}
using the explicit expression of the null BY tensor, as provided in \ref{BYnull}, one obtains
\begin{align}
    \D_a T^a_{~~c}|_{\rm (null)}=\frac{1}{8\pi}\Pi^b_{~c}\Big[\na_a\theta^a_b+\Big(\theta+\kappa\Big)\omega_b+l^a\na_a\omega_b-\Big(\theta+\kappa\Big)k^a\na_al_b-\Pi^a_{~b}\na_a\Big(\theta+\kappa\Big)\Big]\no 
    \\
    -\frac{1}{8\pi}\Pi^b_{~c}\Big(k^i\na_il_a\Big)\Big[\theta^a_b-\Pi^a_{~b}\Big(\theta+\kappa\Big)\Big]~.
\end{align}
Some terms in the above expression cancel each other and the remaining terms can be given as
\begin{align}
    \D_a T^a_{~~c}|_{\rm (null)}=\frac{1}{8\pi}\Pi^b_{~c}\Big[\na_a\theta^a_b+\Big(\theta+\kappa\Big)\omega_b+l^a\na_a\omega_b-\Pi^a_{~b}\na_a\Big(\theta+\kappa\Big)-\theta^a_b(k^i\na_il_a)\Big] \label{APPENNULLCON}
\end{align}
Now, it can be shown that (see Eq. (63) of \cite{Padmanabhan:2010rp})
\begin{align}
    R_{ab}l^b=\na_m\theta^m_a+l^m\na_{m}\omega_a+\Big(\theta+\kappa\Big)\omega_a-\na_a\Big(\theta+\kappa\Big)-\theta_{am}k^{\rm N}\na_nl^m
    \no 
    \\
    -\Big(\omega_m k^{\rm N}\na_nl^m+\na_m k^{\rm N}\na_nl^m+k^{\rm N}\na_m\na_nl^m \Big)l^a~.
\end{align}
This implies 
\begin{align}
    R_{ab}l^b\Pi^a_{~c}=\Pi^a_{~c}\Big[\na_m\theta^m_a+l^m\na_{m}\omega_a+\Big(\theta+\kappa\Big)\omega_a-\na_a\Big(\theta+\kappa\Big)-\theta_{am}k^{\rm N}\na_nl^m\Big]~. \label{identityricci}
\end{align}
Combining \ref{APPENNULLCON} and \ref{identityricci}, we obtain \ref{connullBY}.

\section{Some specific examples} \label{AppC}
\renewcommand{\theequation}{C.\arabic{equation}}
\setcounter{equation}{0}
\subsection{Schwarzschild spacetime}

To illustrate the general formalism, we consider the Schwarzschild spacetime and explicitly compute the null Brown--York tensor and the associated fluid variables.  
In ingoing Eddington--Finkelstein coordinates $(t,r,\psi,\phi)$, the Schwarzschild metric is given by
\begin{align}
ds^2 = -\left(1-\frac{2M}{r}\right)dt^2
+ \frac{4M}{r}\, dt\,dr
+ \left(1+\frac{2M}{r}\right)dr^2
+ r^2\left(d\psi^2+\sin^2\psi\, d\phi^2 \right).
\end{align}

A convenient choice of outgoing null generator and its dual one-form is
\begin{align}
l^a = \left(1,\, \frac{r-2M}{r+2M},\, 0,\, 0 \right),
\qquad
l_a = \left(-\frac{r-2M}{r+2M},\, 1,\, 0,\, 0 \right).
\end{align}
The auxiliary ingoing null vector field is chosen as
\begin{align}
k^a = \left(\frac{2M+r}{2r},\, -\frac{2M+r}{2r},\, 0,\, 0 \right),
\qquad
k_a = \left(-\frac{2M+r}{2r},\, -\frac{2M+r}{2r},\, 0,\, 0 \right).
\end{align}
With this choice, the normalization conditions
\begin{align}
l^a l_a = 0 = k^a k_a,
\qquad
l^a k_a = -1
\end{align}
are easily verified.

The mixed projection tensor $\Pi^{a}{}_{b} = \delta^{a}{}_{b} + k^{a}l_{b}$ takes the form
\begin{align}
\Pi^{a}{}_{b} =
\begin{pmatrix}
\frac{M}{r}+\frac{1}{2} & \frac{M}{r}+\frac{1}{2} & 0 & 0 \\
\frac{1}{2}-\frac{M}{r} & \frac{1}{2}-\frac{M}{r} & 0 & 0 \\
0 & 0 & 1 & 0 \\
0 & 0 & 0 & 1
\end{pmatrix}.
\end{align}
The induced metric on the two-dimensional spatial cross-section $\mathcal B_{(\text{null})}$ is
\begin{align}
q_{ab} =
\begin{pmatrix}
0 & 0 & 0 & 0 \\
0 & 0 & 0 & 0 \\
0 & 0 & r^2 & 0 \\
0 & 0 & 0 & r^2 \sin^2\psi
\end{pmatrix},
\end{align}
which correctly reproduces the metric on a round two-sphere of radius $r$.

The nonvanishing components of the second fundamental form $\theta_{ab}$ are found to be
\begin{align}
\theta_{ab} =
\begin{pmatrix}
0 & 0 & 0 & 0 \\
0 & 0 & 0 & 0 \\
0 & 0 & \dfrac{r(r-2M)}{2M+r} & 0 \\
0 & 0 & 0 & -\dfrac{r\sin^2\psi(2M-r)}{2M+r}
\end{pmatrix}.
\end{align}
Taking the trace with respect to $q^{ab}$ yields the expansion scalar
\begin{align}
\theta = \frac{2(r-2M)}{r(2M+r)}.
\end{align}
As expected for a spherically symmetric spacetime, the shear tensor vanishes identically,
\begin{align}
\sigma_{ab}=0.
\end{align}
The non-affinity parameter (surface gravity) associated with the null generators is
\begin{align}
\kappa = \frac{4M}{(2M+r)^2}.
\end{align}

Using these quantities, the null Brown--York--like energy--momentum tensor defined in \ref{BYnull}  evaluates to
\begin{align}
T^{a}{}_{b}\big|_{\rm (null)} =
\begin{pmatrix}
-\dfrac{r-2M}{8\pi r^2} & -\dfrac{r-2M}{8\pi r^2} & 0 & 0 \\
-\dfrac{(r-2M)^2}{8\pi r^2(2M+r)} & -\dfrac{(r-2M)^2}{8\pi r^2(2M+r)} & 0 & 0 \\
0 & 0 & -\dfrac{-4M^2+4Mr+r^2}{8\pi r(2M+r)^2} & 0 \\
0 & 0 & 0 & -\dfrac{-4M^2+4Mr+r^2}{8\pi r(2M+r)^2}
\end{pmatrix}.
\end{align}

The effective fluid variables on the null surface can now be extracted in a straightforward manner.
The quasilocal energy density is
\begin{align}
\epsilon_{\rm null}
= -T^{a}{}_{b}\big|_{\rm (null)}\, k_a l^b
= \frac{2M-r}{8\pi M r + 4\pi r^2},
\end{align}
which is negative outside the horizon $(r>2M)$, consistent with the attractive nature of gravity.
The momentum density vanishes identically,
\begin{align}
p^c = -T^{a}{}_{b}\big|_{\rm (null)}\, k_a q^{bc} = 0,
\end{align}
reflecting the absence of rotation or angular momentum flux in the Schwarzschild geometry.

Finally, the spatial stress tensor on $\mathcal B_{(\text{null})}$ is given by
\begin{align}
s^{ab} = q^{a}{}_{c} q^{bd} T^{c}{}_{d}\big|_{\rm (null)} =
\begin{pmatrix}
0 & 0 & 0 & 0 \\
0 & 0 & 0 & 0 \\
0 & 0 & -\dfrac{-4M^2+4Mr+r^2}{8\pi r^3(2M+r)^2} & 0 \\
0 & 0 & 0 & -\dfrac{\csc^2\psi(-4M^2+4Mr+r^2)}{8\pi r^3(2M+r)^2}
\end{pmatrix}.
\end{align}
This completes the explicit realization of the null fluid variables for the Schwarzschild spacetime and
provides a concrete consistency check of our general formalism.

\subsection{Kerr spacetime}

We now extend the null Brown--York construction to the Kerr spacetime, describing a stationary, axisymmetric rotating black hole of mass $M$ and angular momentum $J$. In Boyer--Lindquist coordinates $(t,r,\psi,\phi)$, the Kerr metric takes the form
\begin{align}
ds^2 =& -\left(1-\frac{2Mr}{\Sigma}\right)dt^2
+ \frac{\Sigma}{\Delta}\,dr^2
+ \Sigma\, d\psi^2
- \frac{4aMr\sin^2\psi}{\Sigma}\, dt\, d\phi \nonumber\\
&+ \sin^2\psi\left(r^2+a^2+\frac{2a^2Mr\sin^2\psi}{\Sigma}\right)d\phi^2 ,
\end{align}
where
\begin{align}
\Sigma &= r^2 + a^2\cos^2\psi,\\
\Delta &= r^2 - 2Mr + a^2,
\end{align}
and $a \equiv J/M$ denotes the specific angular momentum.

\vspace{0.3cm}

\noindent
To construct the null boundary geometry, we choose a pair of future-directed null vectors $l^a$ and $k^a$, satisfying
$l^a l_a = k^a k_a = 0$ and $l^a k_a = -1$. A convenient choice adapted to the Kerr geometry is
\begin{align}
l^a = \left( \frac{r^2 + a^2}{\Delta},\; 1,\; 0,\; \frac{a}{\Delta} \right),
\qquad
l_a =\left(
-1,\;
\frac{\Sigma}{\Delta},\;
0,\;
a\sin^2\psi
\right),
\end{align}
together with
\begin{align}
k^a=\left(\frac{r^2+a^2}{2 \Sigma},-\frac{\Delta }{2 \Sigma},0,\frac{a}{2 \Sigma}\right),
\qquad
k_a =\left(
-\frac{\Delta}{2\Sigma},\;
-\frac{1}{2},\;
0,\;
\frac{a\,\Delta\,\sin^2\psi}{2\Sigma}
\right).
\end{align}
This choice ensures an affine parametrization of the null generators, as confirmed by the vanishing non-affinity parameter $\kappa=0$.

\vspace{0.3cm}

\noindent
The mixed projector orthogonal to the null pair is defined as
$\Pi^{a}{}_{b} = \delta^{a}{}_{b} + k^a l_b$, which evaluates to
\begin{align}
\Pi^{a}{}_{b}
=
\begin{pmatrix}
1-\dfrac{r^2+a^2}{2\Sigma}
&
\dfrac{r^2+a^2}{2\Delta}
&
0
&
\dfrac{a(r^2+a^2)\sin^2\psi}{2\Sigma}
\\[6pt]
\dfrac{\Delta}{2\Sigma}
&
\dfrac{1}{2}
&
0
&
-\dfrac{a\Delta\sin^2\psi}{2\Sigma}
\\[6pt]
0 & 0 & 1 & 0
\\[6pt]
-\dfrac{a}{2\Sigma}
&
\dfrac{a}{2\Delta}
&
0
&
1+\dfrac{a^2\sin^2\psi}{2\Sigma}
\end{pmatrix}.
\end{align}

The induced metric on the spatial cross-sections of the null hypersurface,
$q_{ab} = g_{ab} + k_a l_b + l_a k_b$, is given by
\begin{align}
q_{ab}
=
\begin{pmatrix}
\dfrac{a^2\sin^2\psi}{\Sigma}
&
0
&
0
&
-\dfrac{a(r^2+a^2)\sin^2\psi}{\Sigma}
\\[6pt]
0 & 0 & 0 & 0
\\[6pt]
0 & 0 & \Sigma & 0
\\[6pt]
-\dfrac{a(r^2+a^2)\sin^2\psi}{\Sigma}
&
0
&
0
&
\dfrac{(r^2+a^2)^2\sin^2\psi}{\Sigma}
\end{pmatrix}.
\end{align}

\vspace{0.3cm}

\noindent
The null extrinsic curvature $\theta_{ab}$, encoding the expansion and shear of the null congruence, is found to be
\begin{align}
\theta_{ab}
=
\begin{pmatrix}
\dfrac{a^2 r \sin^2\psi}{\Sigma^2}
&
0
&
\dfrac{a^2 \sin 2\psi}{2\Sigma}
&
-\dfrac{a r (r^2+a^2)\sin^2\psi}{\Sigma^2}
\\[6pt]
0 & 0 & 0 & 0
\\[6pt]
-\dfrac{a^2 \sin 2\psi}{2\Sigma}
&
0
&
r
&
\dfrac{a (r^2+a^2)\sin 2\psi}{2\Sigma}
\\[6pt]
-\dfrac{a r (r^2+a^2)\sin^2\psi}{\Sigma^2}
&
0
&
-\dfrac{a (r^2+a^2)\sin 2\psi}{2\Sigma}
&
\dfrac{r (r^2+a^2)^2\sin^2\psi}{\Sigma^2}
\end{pmatrix}.
\end{align}
Its trace yields the null expansion
\begin{align}
\theta = \frac{2r}{\Sigma},
\end{align}
We emphasize that the expansion scalar depends on the choice and normalization
of the null generator. The result $\theta = 2r/\Sigma$ reduces to $\theta=2/r$
in the limit $a\to0$, corresponding to the principal null congruence of
Schwarzschild spacetime. This differs from the Eddington--Finkelstein result
obtained earlier, which is adapted to horizon-regular null generators.

The traceless part of $\theta_{ab}$ defines the shear tensor,
\begin{align}
\sigma_{ab}
=
\begin{pmatrix}
0
&
0
&
\dfrac{a^2\sin 2\psi}{2\Sigma}
&
0
\\[6pt]
0 & 0 & 0 & 0
\\[6pt]
-\dfrac{a^2\sin 2\psi}{2\Sigma}
&
0
&
0
&
\dfrac{a(r^2+a^2)\sin 2\psi}{2\Sigma}
\\[6pt]
0
&
0
&
-\dfrac{a(r^2+a^2)\sin 2\psi}{2\Sigma}
&
0
\end{pmatrix},
\end{align}
explicitly capturing anisotropic distortions induced by rotation.

\vspace{0.3cm}

\noindent
The null Brown--York stress tensor on the null boundary is obtained as
\begin{align}
T^{a}{}_{b}\big|_{\rm (null)}
=
\frac{1}{8\pi}
\begin{pmatrix}
-\dfrac{r}{\Sigma}
&
-\dfrac{r(r^2+a^2)}{\Delta\,\Sigma}
&
-\dfrac{a^2\sin 2\psi}{\Sigma}
&
-\dfrac{a r (r^2+a^2)\sin^2\psi}{\Delta\,\Sigma}
\\[8pt]
-\dfrac{r\Delta}{\Sigma^2}
&
-\dfrac{r}{\Sigma}
&
-\dfrac{a^2\sin 2\psi}{2\Sigma}
&
-\dfrac{a r\sin^2\psi}{\Sigma}
\\[8pt]
-\dfrac{a^2\sin 2\psi}{\Sigma^2}
&
0
&
-\dfrac{r}{\Sigma}
&
\dfrac{a(r^2+a^2)\sin 2\psi}{\Sigma^2}
\\[8pt]
\dfrac{a r\sin^2\psi}{\Sigma^2}
&
-\dfrac{a r}{\Delta\,\Sigma}
&
\dfrac{a\cot\psi}{2\Sigma}
&
-\dfrac{r(r^2+a^2)\sin^2\psi}{\Delta\,\Sigma}
\end{pmatrix}.
\end{align}

\vspace{0.3cm}

\noindent
From this tensor, the null fluid variables are readily identified. The energy density measured by the null observer is
\begin{align}
\epsilon_{\rm null}
= -\frac{r}{4\pi\,\Sigma},
\end{align}
which reduces to $\epsilon_{\rm null} = -1/(4\pi r)$ in the limit $a\to0$,
corresponding to the principal null congruence of Schwarzschild spacetime.
The momentum density on the spatial cross-sections is
\begin{align}
p^c =
\left(
-\frac{a^2 r\sin^2\psi}{8\pi\Sigma^2},
\;
0,
\;
-\frac{a^2 \sin 2\psi}{16\pi\Sigma^2},
\;
-\frac{a r}{8\pi\Sigma^2}
\right),
\end{align}
vanishing identically in the non-rotating limit.

Finally, the spatial stress tensor takes the form
\begin{align}
s^{ab} =
\begin{pmatrix}
-\dfrac{a^2 r\sin^2\psi}{8\pi\Sigma^2}
&
0
&
-\dfrac{a^2\sin 2\psi}{16\pi\Sigma^2}
&
-\dfrac{a r}{8\pi\Sigma^2}
\\[8pt]
0 & 0 & 0 & 0
\\[8pt]
\dfrac{a^2\sin 2\psi}{16\pi\Sigma^2}
&
0
&
-\dfrac{r}{8\pi\Sigma^2}
&
\dfrac{a\cot\psi}{8\pi\Sigma^2}
\\[8pt]
-\dfrac{a r}{8\pi\Sigma^2}
&
0
&
-\dfrac{a\cot\psi}{8\pi\Sigma^2}
&
-\dfrac{r\csc^2\psi}{8\pi\Sigma^2}
\end{pmatrix},
\end{align}
demonstrating explicitly how rotation induces anisotropic stresses on the null surface.

In summary, while the Schwarzschild spacetime leads to a remarkably simple null fluid description characterized by vanishing shear, zero momentum density, and isotropic spatial stress, the Kerr spacetime exhibits a much richer structure induced by rotation. The presence of angular momentum introduces non-vanishing shear, anisotropic stresses, and a finite momentum density on the null hypersurface, reflecting the coupling between geometry and frame dragging. Nevertheless, the expansion scalar retains a universal form, reducing smoothly to the Schwarzschild result in the non-rotating limit. This comparison clearly illustrates how rotation imprints itself on the null Brown–York tensor, providing a clean geometric manifestation of rotational degrees of freedom within the null fluid framework.

\subsection{FLRW metric}
Having analyzed stationary black hole spacetimes, it is instructive to briefly
apply our formalism to a cosmological setting. We consider the spatially flat
Friedmann--Robertson--Walker (FRW) spacetime, which describes a homogeneous and
isotropic expanding universe. In spherical polar coordinates, the metric is
given by
\begin{align}
ds^2
= -dt^2
+ a^2(t)\left[
dr^2
+ r^2\left(d\theta^2+\sin^2\theta\,d\phi^2\right)
\right]~,
\end{align}
where $a(t)$ is the cosmological scale factor.
Radial null generators adapted to the cosmological symmetry can be chosen as
\begin{align}
l^a = \left( 1,\; \frac{1}{a(t)},\; 0,\; 0 \right),
\qquad
l_a =\left(
-1,\;
a(t),\;
0,\;
0
\right),
\end{align}
together with the auxiliary null vector
\begin{align}
k^a = \left(\frac{1}{2},\, -\frac{1}{2a(t)},\, 0,\, 0 \right),
\qquad
k_a = \left(-\frac{1}{2},\, -\frac{a(t)}{2},\, 0,\, 0 \right),
\end{align}
which satisfy the normalization conditions
$l^a l_a = 0 = k^a k_a$ and $l^a k_a = -1$.

The mixed projection tensor orthogonal to the null generators is given by
\begin{align}
\Pi^{a}{}_{b}=\left(
\begin{array}{cccc}
 \frac{1}{2} & \frac{a(t)}{2} & 0 & 0 \\
 \frac{1}{2 a(t)} & \frac{1}{2} & 0 & 0 \\
 0 & 0 & 1 & 0 \\
 0 & 0 & 0 & 1
\end{array}
\right),
\end{align}
and the induced metric on the two-dimensional spatial cross-sections
takes the form
\begin{align}
q_{ab}=\left(
\begin{array}{cccc}
 0 & 0 & 0 & 0 \\
 0 & 0 & 0 & 0 \\
 0 & 0 & r^2 a^2(t) & 0 \\
 0 & 0 & 0 & r^2 a^2(t)\sin^2\theta
\end{array}
\right).
\end{align}
As expected, the induced geometry corresponds to a round two-sphere of areal
radius $R=a(t)r$.
The non-vanishing components of the second fundamental form are
\begin{align}
\theta_{ab}=\left(
\begin{array}{cccc}
 0 & 0 & 0 & 0 \\
 0 & 0 & 0 & 0 \\
 0 & 0 & r a(t)\left(r\dot a+1\right) & 0 \\
 0 & 0 & 0 & r a(t)\sin^2\theta\left(r\dot a+1\right)
\end{array}
\right),
\end{align}
leading to the expansion scalar
\begin{align}
\theta
= q^{ab}\theta_{ab}
= \frac{2(r\dot a+1)}{r a(t)}.
\end{align}
The shear tensor vanishes identically,
\begin{align}
\sigma_{ab}=0,
\end{align}
reflecting the homogeneity and isotropy of the FLRW spacetime.

The null Brown--York tensor on the cosmological null boundary is obtained as
\begin{align}
T^{a}{}_{b}\big|_{\rm (null)}
=\left(
\begin{array}{cccc}
 -\frac{r \dot{a}+1}{8 \pi  r a(t)} & -\frac{r \dot{a}+1}{8 \pi  r} & 0 & 0 \\
 -\frac{r \dot{a}+1}{8 \pi  r a(t)^2} & -\frac{r \dot{a}+1}{8 \pi  r a(t)} & 0 & 0 \\
 0 & 0 & -\frac{2 r \dot{a}+1}{8 \pi  r a(t)} & 0 \\
 0 & 0 & 0 & -\frac{2 r \dot{a}+1}{8 \pi  r a(t)}
\end{array}
\right).
\end{align}

The energy density measured by the null observer is therefore
\begin{align}
\epsilon_{\rm null}
= -\frac{1+r\dot a}{4\pi r a(t)},
\end{align}
while the momentum density vanishes, $p^c=0$. The spatial stress tensor is given by
\begin{align}
s^{ab}=\left(
\begin{array}{cccc}
 0 & 0 & 0 & 0 \\
 0 & 0 & 0 & 0 \\
 0 & 0 & -\frac{2 r \dot{a}+1}{8 \pi  r^3 a^3(t)} & 0 \\
 0 & 0 & 0 & -\frac{(2 r \dot{a}+1)\csc^2\theta}{8 \pi  r^3 a^3(t)}
\end{array}
\right).
\end{align}

\section{Energy conservation relation in null hydrodynamics} \label{appenencon}
\renewcommand{\theequation}{D.\arabic{equation}}
\setcounter{equation}{0}
The projection component $R_{ab}l^a l^b$ is related to the dynamical evolution of the expansion scalar corresponding to the null generators $\bml$. This is essentially the null Raychaudhuri equation \cite{Gourgoulhon:2005ng, Poisson:2009pwt}, In the absence of matter, and expanding the L.H.S of Eq. \eqref{connullBY3}, we have the NRE as,
\begin{eqnarray}
	l^a \nabla_a \th - \k \th + \fr \th^2 + \sigma_{ab}\sigma^{ab} = 0~.
	\label{manip1}
\end{eqnarray}
Since the evolution of the expansion scalar is considered along the null generators of $\hech$, we have in the adapted coordinate system, $l^{\mu} \overset{\hech}{=} (1, V^A)$. Thus expanding $l^a \nabla_a \th = \p_t \th  + V^A \p_A \th$, we have from  \eqref{manip1},
\begin{eqnarray}
	-\p_t \th - v^A \p_A \th + \k \th - \th^2 = \sigma_{ab} \sigma^{ab} - \fr \th^2 ~.
	\label{manip2}
\end{eqnarray}
Now, we consider the specific case in which the induced metric on the submanifold $\sptqr$ is stationary or independent of the time evolution parameter $t$, i.e. $\p_t q_{AB} \overset{\hech}{=} 0$. Then from \eqref{manip34}, we have that the second fundamental form takes the structure,
\begin{eqnarray}
	\theta_{AB} \overset{\hech}{=} \fr\Big(\twod_A V_B + \twod_B V_A\Big) ~,
	\label{manip3}
\end{eqnarray}
with the expansion scalar being given as $\th \overset{\hech}{=} \twod_A V^A$. Inserting this in \eqref{manip2}, we have,
	\begin{eqnarray}
		\p_t (-\th) + V^A \p_A (-\th) + (\k - \th)\twod_A V^A = \sigma_{ab}\sigma^{ab}  - \fr (\twod_A V^A )^2~.
		\label{manip4}
	\end{eqnarray}
Dividing \eqref{manip4} by a factor of $1/8 \pi$ essentially leads us to \eqref{energyeqn}.

\bibliographystyle{elsarticle-num}
\bibliography{SB-reference1x}

\end{document}